%% file: Accumulator.tex
\documentclass[prd,longbibliography,nofootinbib,twocolumn,preprintnumbers]{revtex4-1}
\input{packages}
\input{macros}

\usepackage{soul}

\begin{document}

\preprint{FERMILAB-PUB-25-0622-SQMS-T}

\title{Electric Accumulation of Millicharged Particles}

\author{Asher Berlin$^{a,b}$}
\email{aberlin@fnal.gov}
\author{Zachary Bogorad$^{a,b}$}
\email{zbogorad@fnal.gov}
\author{Peter W.~Graham$^{c,d}$}
\email{pwgraham@stanford.edu}
\author{Harikrishnan Ramani$^{e}$}
\email{hramani@udel.edu}
\affiliation{$^a$Theory Division, Fermi National Accelerator Laboratory}
\affiliation{$^b$Superconducting Quantum Materials and Systems Center (SQMS), Fermi National Accelerator Laboratory}
\affiliation{$^c$Leinweber Institute for Theoretical Physics at Stanford, Department of Physics, Stanford University}
\affiliation{$^d$Kavli Institute for Particle Astrophysics \& Cosmology, Department of Physics, Stanford
University}
\affiliation{$^e$Department of Physics and Astronomy, University of Delaware and the Bartol Research Institute}

\begin{abstract}
A terrestrial population of millicharged particles that interact significantly with normal matter can arise if they make up a dark matter subcomponent or if they are light enough to be produced in cosmic ray air showers. Such particles thermalize to terrestrial temperatures through repeated scatters with normal matter in Earth's environment. We show that a simple electrified shell (e.g., a Van de Graaff generator) functions as an efficient accumulator of such room-temperature millicharged particles, parametrically enhancing their local density by as much as twelve orders of magnitude. This can be used to boost the sensitivity of any detector housed in the shell's interior, such as ion traps and tests of Coulomb's law. In a companion paper, we apply this specifically to Cavendish tests of Coulomb's law, and show that a well-established setup can probe a large region of unexplored parameter space, with sensitivity to the irreducible density of millicharged particles generated from cosmic rays that outperforms future accelerator searches for sub-GeV masses. 
\end{abstract}

\maketitle

{\hypersetup{linkcolor=black}
\tableofcontents}

\section{Introduction}
\label{sec:intro}

Although many theories have been proposed to solve the plethora of outstanding puzzles indicating that the Standard Model (SM) is incomplete, the new particles predicted by these theories have yet to be directly observed. It is widely believed that this is a consequence of these particles possessing either a feeble coupling beyond the sensitivity of precision experiments  or a mass above the reach of high-energy colliders.   

There is, however, a simple and largely unconstrained extension to the SM that does not involve extremely small interactions or large masses. These are theories that predict new particles $\x$ with a small effective electromagnetic charge $q_\x \lesssim 10^{-1}$, referred to as millicharged particles (mCPs).
Such particles can naturally arise from the kinetic mixing between a new long-ranged $\Uone$ gauge boson and the SM photon~\cite{Holdom:1985ag}. They also are viable dark matter (DM) candidates and have been invoked to explain a multitude of experimental anomalies~\cite{PVLAS:2005sku,Chang:2008aa,PAMELA:2008gwm,Barkana:2018lgd,Berlin:2018sjs,Barkana:2018qrx,Liu:2019knx}. This minimal extension to the SM is largely unexplored despite decades of scrutiny, with viable parameter space remaining for, e.g., charges as large as $q_\x \sim 10^{-1}$ for GeV-scale masses.

For a large range of couplings and masses, mCPs are strongly-coupled to the SM due to the long-ranged nature of the interaction. For instance, if present on Earth, these particles can rapidly scatter off of terrestrial matter, thereby thermalizing down to ambient temperatures and building up to non-negligible densities~\cite{Pospelov:2020ktu,Berlin:2023zpn}. Such a terrestrial population can arise if, e.g., mCPs make up a small subcomponent of the total Galactic DM density, or if they are light enough to be produced on Earth in nuclear decays~\cite{Gao:2025ykc} or cosmic ray air showers~\cite{Plestid:2020kdm}. Note that in the latter case, a search for this irreducible population is equivalent to testing the model itself, akin to an accelerator search, and does not require the mCPs to make up any of the DM.

Since ambient temperatures of $300 \ \text{K} \sim 25 \ \meV$ are well below the energy thresholds of typical experimental sensors, a dense terrestrial population of strongly-coupled mCPs remains an untested possibility. Various detection strategies have been investigated~\cite{Carney:2021irt,Budker:2021quh,Afek:2020lek,Berlin:2023gvx}, with the strongest current limits  arising from considerations of anomalous heating in ion trap experiments~\cite{Carney:2021irt,Budker:2021quh}.

\begin{figure}[t!]
\centering
\includegraphics[width=0.47 \textwidth]{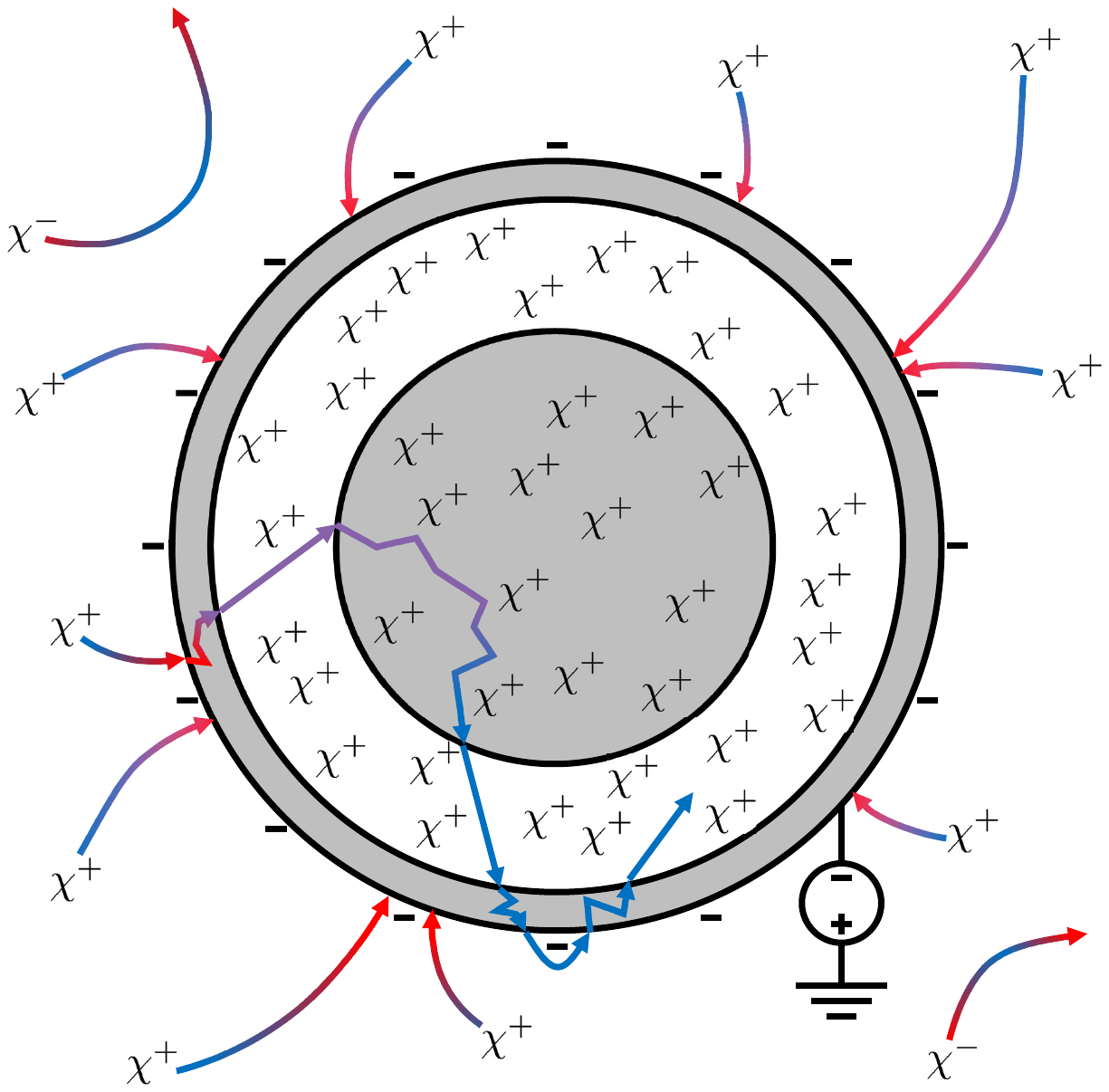}
\caption{A schematic of our experimental proposal. A large ``accumulator shell" of radius $R_0$ is charged to strong negative voltage, $\phi_0$. This attracts positively-charged millicharged particles $\x^+$, which lose energy after scattering on an enclosed solid sphere, and become electrically trapped within the cavity of the shell, parametrically enhancing their local density. The trajectories of various millicharged particles are shown with arrows, where the color of the arrow is meant to show regions where the particles are moving with greater (red) or smaller (blue) kinetic energy. A detector can be placed inside the shell to detect this millicharge overdensity (see Ref.~\cite{forthcoming}). }
\label{fig:cartoon}
\end{figure}

In this work, we propose a simple modification that drastically increases the sensitivity of any of these experimental strategies. In particular, the local density of mCPs (irrespective of their origin) can be enhanced by enclosing a  detector within a shell of fixed voltage. As shown schematically in \Fig{cartoon}, this electrified shell functions as a trap by dragging ambient mCPs inwards where they scatter, become electrically bound to the interior region, and accumulate to much larger densities. We find that such an \emph{accumulator} operating for a time $t$ can  enhance the local density of mCPs by as much as $\sim 10^{12} \times (t / \yr)$ compared to the average terrestrial value. 

Since the interior of the accumulator experiences negligible electric fields (aside from that sourced by mCPs), it can, in principle, house a precision detector sensitive to the induced millicharge overdensity. While we focus on the general implementation of an accumulator in this work, we note that it can enable various search strategies to probe a large region of unexplored parameter space. In a companion paper~\cite{forthcoming}, we apply this to a simple experimental setup with the goal of measuring the small electric field sourced by the millicharge overdensity. As shown there, an accumulator paired with decades-old detection technology can enable sensitivity to the irreducible mCP density generated from cosmic rays, at a level which outperforms future accelerator searches for sub-GeV masses. 

The rest of this paper is organized as follows. In \Sec{model}, we provide an overview of mCP models. In \Sec{overdensity}, we discuss how such particles behave on Earth, and calculate the irreducible density of mCPs sourced by cosmic rays. In \Sec{compactor}, we discuss how mCPs accumulate within an electrified shell. We present our final results in \Sec{results}, and conclude in \Sec{conclusion}. Appendices are referred to throughout the main text, which provide additional technical details of our analysis.

\section{Model Overview}
\label{sec:model}

%

Particles with small effective electromagnetic charge arise naturally within the context of dark sectors charged under a new dark photon $\Ap$ that kinetically-mixes with the SM photon~\cite{Holdom:1985ag},
\be
\Lag \supset \frac{\eps}{2} \, F_{\mn}^\p \, F^{\mn} - j_\mu^\p \, A^{\p \, \mu}
~,
\ee
where $F_{\mn}^\p$ and $F_{\mn}$ are the dark and SM field-strengths, $\eps$ is the small dimensionless kinetic-mixing parameter, and $j_\mu^\p$ is the dark current arising from particles $\x$ directly charged under the $\Ap$. 

The kinetic-mixing term can be diagonalized to leading order in $\eps \ll 1$ by the field-redefinition $\Ap \to \Ap + \eps \, A$, where $A$ is the SM photon field. As a result, $\x$ picks up an effective charge under the SM photon of $e q_\x \simeq \eps \, e^\p$, where $e^\p$ is the dark gauge coupling. We note that similar conclusions hold if the dark photon has a non-zero mass $\mAp$. However, in this case $\x$ behaves as though it has an effective electromagnetic charge only on length scales smaller than the dark photon's Compton wavelength $\sim \mAp^{-1}$, since at greater distances any long-ranged interaction mediated by the $\Ap$ is exponentially screened~\cite{Berlin:2019uco,Berlin:2023zpn}. In this work, we will take the interaction to be long-ranged on laboratory scales, corresponding to $\mAp \lesssim 1 \ \m^{-1} \sim 10^{-7} \ \eV$, and will comment on the impact of particular values of $\mAp$ when relevant. 

In the simplest models, $\eps$ is generated by loops of $N$ generations of heavy particles charged under both sectors, such that $\eps \sim N e^\p e / (4 \pi)^2$ and hence $q_\x \sim N \alpha^\p / 4 \pi$, where $\alpha^\p = e^{\p \, 2} / 4 \pi$ is the dark fine-structure constant. It is therefore natural to expect millicharges of size $q_\x \sim N \times 10^{-3}$ for a dark gauge coupling comparable to ones in the SM. Millicharges parametrically smaller than this value are also possible in scenarios where $e^\p \ll e$, or those in which $\eps$ is generated to leading order at two-loops or via higher-dimensional operators~\cite{Arkani-Hamed:2008kxc,Gherghetta:2019coi}.

Alternatively, $\x$ may couple directly to the SM photon if it carries an electric charge under electromagnetism~\cite{wen1985electric,gell1978color,athanasiu1988remarks,Langacker:2011db}. In this scenario, gauge transformations under the SM $\Uone$ are well-defined provided that both the mCP and electron charges, $q_\x$ and $q_e$, take integer values with respect to some fundamental charge unit, which may be much smaller than the electron charge. 
Unlike models with massive dark photons, here the interaction is arbitrarily long-ranged in vacuum. When relevant, we will discuss how our analysis depends on these model-specific details.


\section{Terrestrial Population}
\label{sec:overdensity}

%
\begin{figure}[t!]
\centering
\includegraphics[width=0.495 \textwidth]{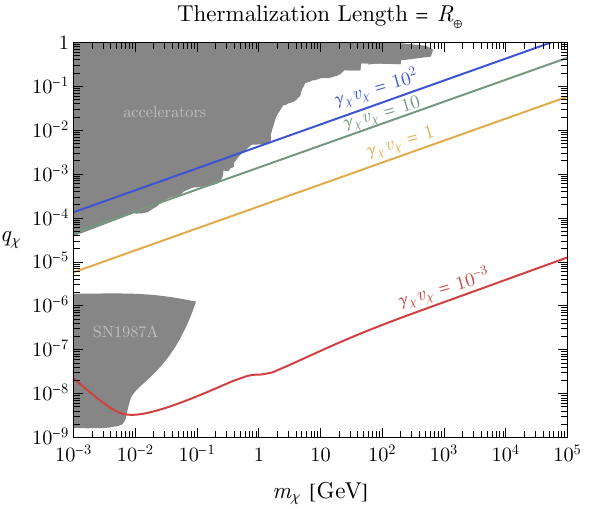}
\caption{The minimum coupling $q_\x$ required for a millicharged particle of mass $m_\x$ to cool down to room temperature within an Earth radius after scattering with terrestrial atoms, for various choices of the particle's initial boost times velocity $\g_\x v_\x$ ($\g_\x v_\x = 10^{-3}$ for millicharged dark matter, and $\g_\x v_\x \gtrsim 1$ for relativistic millicharged particles, such as those created by cosmic rays). In gray, we also show existing limits from accelerator probes~\cite{Davidson:2000hf,Haas:2014dda,Prinz:1998ua,ArgoNeuT:2019ckq,milliQan:2021lne,ArguellesDelgado:2021lek,PBC:2025sny,CMS:2024eyx,Alcott:2025rxn} and SN1987A~\cite{Chang:2018rso}.}
\label{fig:penetration}
\end{figure}

We now provide a brief overview of how a terrestrial mCP population can arise. For a more detailed discussion, see, e.g., Refs.~\cite{Pospelov:2020ktu,Berlin:2023zpn}. 

For sufficiently large charges, mCPs that pass through the Earth drastically slow down due to collisions with terrestrial atoms. The minimum coupling required for this to occur within an Earth radius $R_\oplus$ is shown in \Fig{penetration} for various choices of the mCP's initial boost times velocity $\g_\x v_\x$. For $\g_\x v_\x \gg 1$, we follow the standard analytic estimates for stopping power via ionizing scatters (as in, e.g.,  Ref.~\cite{ArguellesDelgado:2021lek}), while for $\g_\x v_\x \ll 1$, we adopt the calculation from Ref.~\cite{Berlin:2023zpn} for energy loss through elastic atomic scatters. For intermediate values $\g_\x v_\x \sim 1$, we adopt the results from the data-driven analysis in Ref.~\cite{Harnik:2020ugb}.

For couplings $q_\x \gtrsim m_e / \mu_{\x N}$, negatively-charged mCPs are efficiently captured into bound states with atomic nuclei, where $m_e$ is the electron mass and $\mu_{\x N}$ is the mCP-nuclear reduced mass~\cite{Pospelov:2020ktu,Berlin:2023zpn}. We note that calculations in \Secs{compactor}{results} partially extend into this region of parameter space. However, since our treatment  assumes that mCPs are unbound, our results should be applied only to positively-charged mCPs, which exist as a free plasma on Earth throughout most of our parameter space of interest; since even this ceases to be true if $q_\x \gtrsim \sqrt{(300 \ \text{K} ~ m_e ) / (\text{Ry} \, \mu_{\x e})} \simeq 4 \times 10^{-2}$ for $m_\x \gg m_e$ (in which case  positively-charged mCPs form room-temperature bound states with electrons),  we refrain from considering such charges entirely. We also note that various searches limit the abundance of mCPs bound to terrestrial matter~\cite{Marinelli:1982dg,Kim:2007zzs,Moore:2014yba,Afek:2020lek}, but these typically apply to larger couplings or larger ambient densities than we consider here.

For $m_\x \gg 1 \ \GeV$, these mCPs become gravitationally bound and accumulate on Earth after slowing down, leading to large overdensities~\cite{Pospelov:2020ktu,Berlin:2023zpn}. Conversely, for $m_\x \lesssim 1 \ \GeV$, their thermal velocity at room temperature exceeds the terrestrial escape velocity. Thus, absent any additional interactions, such particles evaporate off of the Earth. Even in this case, however, the increased residence time spent by mCPs as they slowly diffuse throughout the Earth before evaporating can lead to a large overdensity~\cite{Berlin:2023zpn}. 

For any mass, the terrestrial density of positively-charged mCPs can be further enhanced by the trapping effects of the atmospheric electric field~\cite{Berlin:2023zpn}, which possesses a typical value of $E_\oplus \sim 1 \ \V / \cm$ near the surface. $E_\oplus$ exists throughout the atmosphere, giving rise to an  average voltage difference $\phi_\oplus \sim 0.3 ~ \text{MV}$ between the ionosphere and crust. This value of $\phi_\oplus$ is the typical one during fair weather conditions and is maintained by the global atmospheric electrical circuit. During thunderstorms or geomagnetic activity, the potential profile of $\phi_\oplus$ is locally distorted on small scales, but is otherwise stable between the crust and ionosphere. We thus do not expect weather activity to change these conclusions, although a detailed study is beyond the scope of this work. 

Therefore, positively-charged mCPs with $q_\x \gtrsim (300 \ \text{K}) / (0.3 \ \MeV) \sim 10^{-7}$  are trapped by $\phi_\oplus$ after thermalizing to room temperature and reside primarily below the crust, extending a height $\sim 300 \ \text{K} / (e q_\x \, E_\oplus) \sim 1 \ \mm \times q_\x^{-1}$ above the surface. However, for interactions mediated by a kinetically-mixed dark photon, this electrical trapping saturates once the accumulation of positively-charged mCPs sources a significantly strong repulsive \emph{dark} electric field to overcome the effect of Earth's visible field, corresponding to a volume-averaged density of
\be
\label{eq:maxCR}
n_\x \sim \frac{3 e q_\x \, E_\oplus}{e^{\p \, 2} \, R_\oplus} 
\sim 10^{-5} \ \cm^{-3} \times \bigg( \frac{q_\x}{10^{-3}} \bigg) \, \bigg( \frac{10^{-1}}{e^\p} \bigg)^2
~.
\ee
For models involving kinetically-mixed dark photons, \Eq{maxCR} corresponds to the maximal value of the mCP density (averaged over the Earth) that can be bound by the atmospheric electric field. Strictly speaking, this is a bound on the difference between the positive and negative mCPs ($n_{\x^+}-n_{\x^-}$), and hence $n_{\x^+} + n_{\x^-}$ can exceed this upper limit. However, in this case, there is no benefit from the trapping effect of $E_\oplus$.

For kinetically-mixed models, the self-repulsive force from this dark electric field repels the mCPs to the Earth's surface before the density saturates to the average value shown in \Eq{maxCR} (this effect vanishes for mCPs directly coupled to the photon due to the conductivity of the Earth). This occurs once the dark repulsion overcomes the mCP's thermal kinetic energy as well as Earth's gravitational field $g_\oplus$, corresponding to volume-averaged densities of
\be
\label{eq:darkEoverT}
n_\x \sim \frac{300 \ \text{K}}{(e^\p \, R_\oplus)^2} \sim 10^{-12} \ \cm^{-3} 
\times \bigg( \frac{10^{-1}}{e^\p} \bigg)^2
~,
\ee
and
\be
\label{eq:darkEoverg}
n_\x \sim \frac{3 m_\x \, g_\oplus}{e^{\p \, 2} \, R_\oplus} \sim 10^{-11} \ \cm^{-3} 
\times \bigg( \frac{m_\x}{\GeV}\bigg) \, \bigg( \frac{10^{-1}}{e^\p} \bigg)^2
~,
\ee
respectively. Once the average mCP density (more precisely the asymmetry $n_{\x^+} - n_{\x^-}$) exceeds both of these values, this population of mCPs will reside primarily near the surface. More specifically, energy conservation (i.e., that the total thermal kinetic energy of the mCPs is comparable to the work required to change the radius occupied by this population) implies that such particles  occupy an annulus of radius $R_\oplus$ and radial thickness $\delta r_\x \sim 300 \ \text{K} \, / \, (e^\p \, \langle n_\x \rangle \, R_\oplus)$, where $\langle n_\x \rangle$ is the mCP density volume-averaged over the Earth. As a result, the local density near the surface is significantly enhanced compared to the volume-averaged one by $R_\oplus / \delta r_\x \sim \langle m_D^{\p} \rangle^2 \, R_\oplus^2 \gg 1$, where $\langle m_D^\p \rangle$ is the mCP's contribution to the volume-averaged Debye mass of the dark photon. 

Furthermore, the symmetric abundance of mCPs can be affected by mCP annihilations into electrons or dark photons, provided that negatively-charged mCPs do not efficiently bind to nuclei. However, this depends on the particular size of the terrestrial density, and is generally not relevant for the populations that we consider in this work.

In the subsections below, we apply the above considerations to two example sources of a terrestrial overdensity of mCPs. This corresponds to mCPs that make up a DM subcomponent, and the irreducible density that is produced in cosmic ray air showers. Below, we focus primarily on mCPs that couple directly to the SM photon since models involving kinetic mixing introduce additional dynamics that can significantly enhance or suppress the terrestrial density, depending on the particular value of the dark photon mass and gauge coupling. Therefore, the results shown in the figures below apply predominantly to direct mCP-photon interactions; modifications that arise in models involving kinetically-mixed dark photons will be explicitly discussed when relevant.

\subsection{Dark Matter Subcomponent}
\label{sec:mCPDM}

For $q_\x \gtrsim 10^{-7} \times (m_\x / \GeV)^{1/2}$, corresponding to most of the parameter space of interest, mCPs equilibrate with the SM bath in the early universe when the radiation temperature is $T \gtrsim m_\x$. These large couplings also imply that mCPs remain tightly-coupled to the photon-baryon fluid during the epoch of recombination, which is highly constrained by measurements of the cosmic microwave background if mCPs constitute an $\order{1}$ fraction of the DM density. However, cosmological data does not preclude the possibility that a sub-percent DM subcomponent interacts strongly with the SM~\cite{Dubovsky:2003yn,dePutter:2018xte,Kovetz:2018zan,Buen-Abad:2021mvc}.

Thus, the thermal cosmological abundance of strongly-coupled mCPs must be depleted to a small fraction $f_\DM \ll 1$ of the total DM density. This can occur if, e.g., they efficiently annihilate to pairs of SM fermions or dark photons, which is possible for $q_\x \gg 10^{-3} \times (m_\x / \GeV)$ or $e^\p \gg 10^{-2} \times (m_\x / \GeV)^{1/2}$, respectively. More generally, mCP annihilations to lighter states that saturate perturbative unitarity can reduce the thermal abundance to fractional densities as small as $f_\DM \sim 10^{-10} \times (m_\x / \GeV)^2$~\cite{Griest:1989wd}. This fraction can be further reduced to exponentially smaller values if, for instance, the initial reheat temperature of the universe is significantly less than $m_\x$, in which case mCPs do not efficiently thermalize with the SM bath~\cite{Berlin:2021zbv,Gan:2023jbs,Boddy:2024vgt,Bernal:2024ndy}.

Strongly-coupled relics with $f_\DM \ll 1$ are also relatively unconstrained by terrestrial direct detection experiments searching for DM scattering. For instance, these experiments are signal-limited for DM fractions of $f_\DM \lesssim 10^{-8}$~\cite{Pospelov:2020ktu}. Typical experiments are also insensitive to any value of $f_\DM$ if $q_\x \gtrsim 10^{-4} \times (m_\x / \GeV)^{1/2}$, in which case mCP DM scatters and rapidly cools down to room temperature, well below threshold, before encountering near-surface detectors. A detailed analysis of the sensitivity of direct detection experiments in this strongly-coupled regime was performed in Ref.~\cite{Emken:2019tni}, which ignored model-dependent long-distance effects arising from the solar wind and terrestrial magnetic fields.

Upon cooling down to room temperature, the local phase space of mCP DM is drastically modified compared to the Galactic population. As shown in \Fig{penetration}, an incoming millicharged DM particle with velocity $v_\x \sim 10^{-3}$ and charge $q_\x \gtrsim \text{few} \times 10^{-8} \times (m_\x / \GeV)^{1/2}$ scatters with terrestrial matter and sheds an $\order{1}$ fraction of its kinetic energy on Earth~\cite{Berlin:2023zpn}. By conservation of flux, this leads to large local overdensities. For $m_\x \gg 1 \ \GeV$, these particles are gravitationally bound and accumulate on Earth over terrestrial timescales to densities as large as $\sim 10^{16}$ times their Galactic density for a large range of couplings~\cite{Berlin:2023zpn}. For $m_\x \ll 1 \ \GeV$, their increased residence time spent on Earth before evaporation can result in overdensities as large as $\sim 10^8$~\cite{Berlin:2023zpn}.

The predicted size of these overdensities can be significantly modified if $\x$ couples to electromagnetic fields on planetary or solar length-scales, which in turn depends on the particular value of the $\Ap$ mass~\cite{Dunsky:2018mqs,Emken:2019tni,Berlin:2023zpn}. For instance, for $\mAp \lesssim R_\odot^{-1} \sim 10^{-16} \ \eV$ or $\mAp \lesssim R_\oplus^{-1} \sim 10^{-14} \ \eV$ (where $R_\odot$ and $R_\oplus$ are the solar and Earth radius, respectively), mCPs can efficiently couple to the long-ranged magnetic field generated by the solar wind or the Earth, reducing their ability to penetrate into Earth's environment. For $\mAp \lesssim h_\text{atm}^{-1} \sim 10^{-12} \ \eV$ (where $h_\text{atm}$ is the height of the ionosphere) and $q_\x \gg 10^{-7}$, the coupling of mCPs to the atmospheric electric field can instead enhance the trapping of positively-charged mCPs below the atmosphere, as discussed in detail above.

Due to the strong model-dependence of these long-ranged electromagnetic effects, we remain agnostic regarding their importance for mCP DM by parameterizing the abundance of mCP DM solely in terms of its local terrestrial number density, denoted as $n_\x$. Translating this to, e.g., the mCP's Galactic density requires specification of the $\Ap$ mass and detailed modeling of Galactic, solar, and terrestrial electromagnetic fields, which is beyond the scope of this work.

\subsection{Irreducible Cosmic Ray Population}
\label{sec:CRpop}

Independent of the Galactic population, an irreducible density of terrestrial mCPs arises from cosmic ray air showers. In particular, for $m_\x \ll 10 \ \GeV$, the dominant production mechanism is through the decay of secondary cosmic ray mesons. The salient feature of this production mechanism is that it is independent of model-dependent details regarding the early universe and propagation through the solar and terrestrial magnetic fields, since such particles are produced locally  in Earth's atmosphere. 

The relativistic population of cosmic ray mCPs has been worked out in detail in, e.g., Refs.~\cite{Plestid:2020kdm,Harnik:2020ugb,ArguellesDelgado:2021lek,Du:2022hms,Wu:2024iqm}, where it was shown that neutrino and DM detectors have some sensitivity to these particles, ultimately limited by the small flux. Although most of these mCPs are produced relativistically, they can efficiently scatter in Earth's crust, thermalizing to terrestrial temperatures. Even if the resulting non-relativistic population eventually evaporates from the Earth, its slow diffusion throughout the atmosphere implies an enhancement to the mCP density.

In this work, we use the results of Ref.~\cite{Berlin:2023zpn}, which calculated this \emph{non-relativistic} population of mCPs produced by cosmic rays. For the sake of brevity, we simply quote here the main relevant results. If mCPs are not bound by Earth's gravitational or electric fields, then they eventually evaporate. In this case, the resulting density of non-relativistic mCPs at a depth $x > 0$ below the top of the atmosphere is~\cite{Berlin:2023zpn}
\be
\label{eq:trafficjam}
n_\x (x) \simeq 
\int d E_\x \,  \frac{d \Phi_\x^{(\text{CR})}}{d E_\x}  \, \text{Pr}_\text{th}^\oplus \, \frac{\min{(x , x_\text{th})} - x_\text{LSS}}{D_\x}
 ~,
\ee
where $x_\text{th} > 0$ is the depth below the top of the atmosphere at which an initially-produced mCP cools down to ambient temperature through repeated scattering, and $x_\text{LSS} > 0$ is the depth of the ``last scattering surface," i.e., the point at which an outgoing thermalized mCP stops scattering and free-streams away from the Earth. Above, the integral is over the initial energy $E_\x$ of the mCP, $\Phi_\x^\text{(CR)}$ is the flux of mCPs generated by cosmic rays,  $\text{Pr}_\text{th}^\oplus$ is the probability that an mCP cools down to ambient temperature in Earth, and $D_\x$ is the diffusion coefficient of the thermalized mCP (parametrically the product of the collisional mean free path and thermal velocity). 

Millicharged particles can instead accumulate on Earth if they become gravitationally or electromagnetically bound. In the former case, this is possible for $m_\x \gg 1 \ \GeV$. However, cosmic-ray production of particles much heavier than a GeV is rare, and so this applies only to a small part of parameter space. It is thus more generic for mCPs produced by cosmic rays to become electromagnetically bound to the Earth~\cite{Berlin:2023zpn}. As discussed above, the degree to which this can occur depends on the nature of the interaction. Unlike mCP DM, however, mCPs produced by cosmic rays are initially relativistic, such that they can easily enter Earth's environment, independent of their couplings to long-ranged magnetic fields. 

\begin{figure*}[t!]
\centering
\includegraphics[width=0.495 \textwidth]{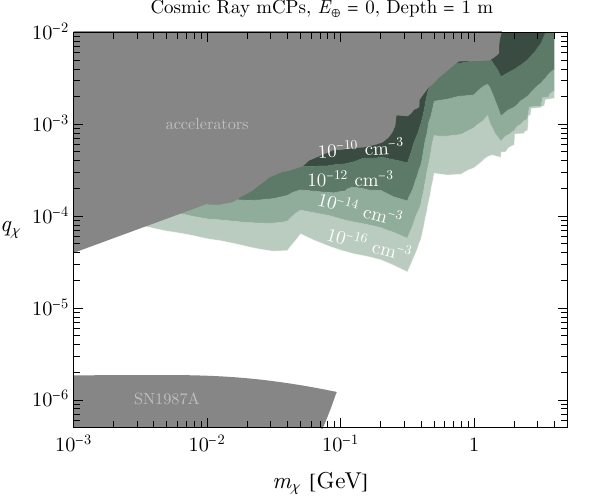}
\includegraphics[width=0.495 \textwidth]{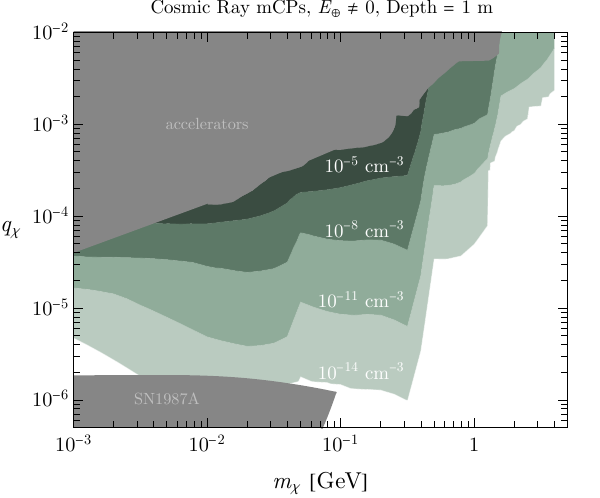}
\caption{The terrestrial density $n_\x$ of positively-charged millicharged particles that are produced by cosmic rays and thermalize to room temperature through scattering with Earth's environment, as a function of the particle's mass $m_\x$ and charge $q_\x$. In gray, we also show existing limits from accelerator probes~\cite{Davidson:2000hf,Haas:2014dda,Prinz:1998ua,ArgoNeuT:2019ckq,milliQan:2021lne,ArguellesDelgado:2021lek,PBC:2025sny,CMS:2024eyx,Alcott:2025rxn} and SN1987A~\cite{Chang:2018rso}. In the left-panel, ``$E_\oplus = 0$" means we ignore the coupling of millicharged particles to the atmospheric electric field. In the right-panel, ``$E_\oplus \neq 0$" means we instead assume that the millicharged particles couple to the atmospheric electric field, such that positively-charged particles can become electrically bound below the atmosphere. In both panels, we evaluate the density at a depth of $1 \ \m$ below the crust (these results also apply to the density inside of a conducting building  above the surface). In the left-panel, there is substantial depth-dependence to $n_\x$, with greater densities further underground. In the right-panel, there is negligible depth-dependence below the crust.
}
\label{fig:nxCR}
\end{figure*}

After such mCPs scatter and thermalize to room temperature, their small energy can suppress their ability to overcome Earth's electromagnetic fields. For instance, for $q_\x \gtrsim 10^{-9} \times \sqrt{m_\x / \MeV}$, their gyroradius in Earth's magnetic field is much smaller than $R_\oplus$, meaning that such particles remain confined except possibly  near the poles. More importantly, as discussed above, positively-charged mCPs remain bound by the atmospheric electric field for $q_\x \gtrsim 10^{-7}$. However, due to modifications that can arise in kinetically-mixed models, we will instead consider the two extremes where mCPs do or do not become trapped by Earth's electromagnetic fields over the age of the Earth, $t_\oplus \simeq 4.5 \times 10^9 \ \yr$.\footnote{Direct evidence in the form of fulgurites shows that the global atmospheric electrical circuit has been active for at least $\sim (0.25 - 0.5) \times 10^9 \ \yr$, while indirect evidence points to much longer timescales, comparable to $t_\oplus$~\cite{Harrison2004}.} 

In the case that thermalized mCPs produced by cosmic rays become gravitationally or electromagnetically bound, they accumulate for a time $t_\oplus$. The density of this population, volume-averaged over the Earth, is given by~\cite{Berlin:2023zpn}
\be
n_\x \simeq \frac{3 \, t_\oplus}{R_\oplus} \, \int d E_\x \,  \frac{d \Phi_\x^{(\text{CR})}}{d E_\x}  \, \text{Pr}_\text{th}^\oplus
~.
\ee
The radial profile of this density depends on the nature of the mCP interaction. For instance, if mCPs directly couple to the SM photon, then their density is approximately uniform below the crust for $m_\x \lesssim \text{few} \times \GeV$ or highly-peaked near Earth's center for $m_\x \gg \text{few} \times \GeV$ (due to Earth's gravitational field)~\cite{Berlin:2023zpn}. Instead, if mCPs interact through a kinetically-mixed dark photon and become electromagnetically bound by Earth's atmospheric voltage, the repulsive effect from the resulting dark electric field can cause mCPs to reside primarily near the surface, parametrically enhancing their local density (see the discussion near \Eqs{darkEoverg}{darkEoverT}). Due to the significant model-dependence of this effect, we will focus on mCPs that couple directly to the SM photon in our results below.

Using the full formalism outlined in Ref.~\cite{Berlin:2023zpn}, we determine the thermalized mCP density produced by cosmic rays. We improve upon the analysis of Ref.~\cite{Berlin:2023zpn}, however, by making use of the publicly available code of Ref.~\cite{ArguellesDelgado:2021lek} for determining $\Phi_\x^\text{(CR)}$, which includes the dominant contributions for sub-GeV mCPs arising from decays of $\pi$, $\eta$, $\rho$,  $\omega$, $\phi$, and $J/\psi$ mesons. We have also supplemented this with $\Upsilon$ decays, following the semianalytic procedure of Ref.~\cite{Wu:2024iqm}. Although we do not include other processes, we note that for masses $m_\x \gtrsim 10 \ \GeV$, Drell-Yan dominates the total yield~\cite{Wu:2024iqm}.

It has also recently been claimed in Refs.~\cite{Du:2022hms,Wu:2024iqm} that proton bremsstrahlung is the dominant production mechanism for mCPs in the $(0.1 - 1) \ \GeV$ mass range, since it benefits from the resonant enhancement of photon-vector meson mixing. However, this resonant enhancement corresponds to one subprocess of vector meson production. Since the latter has already been incorporated in our analysis, as well as in other studies, it remains unclear as to why bremsstrahlung is found to yield significantly larger fluxes. For this reason, we conservatively choose to not incorporate bremsstrahlung into our analysis. 

Our results for the irreducible non-relativistic density of mCPs sourced by cosmic rays is shown in \Fig{nxCR}, evaluated at a depth of $1 \ \m$ below the surface of the crust (these results also apply to the density inside of a conducting building  above the surface). In the left-panel, we assume that mCPs do not couple to the atmospheric electric field, such that they do not remain bound to the Earth after thermalizing to room temperature. In this case, there is substantial depth-dependence to $n_\x$ with greater densities further underground. This is because such mCPs are not bound to the Earth and thermalize to room temperature very deep ($\gg 1 \ \km$) underground, leading to the approximately-linear scaling with depth in \Eq{trafficjam}. 

In the right-panel, we instead take the mCPs to couple directly to the atmospheric electric field $E_\oplus \sim 1 \ \V / \cm$, such that they are trapped below the atmosphere for $q_\x \gg 10^{-7}$. This drastically enhances the terrestrial density of sub-GeV mCPs that would otherwise evaporate from the Earth and results in densities as large as $n_\x \sim 10^{-5} \ \cm^{-3} \times (q_\x / 10^{-4})^2 \, $ for $m_\x \ll 1 \ \GeV$. In this case, such mCPs remain bound to the Earth, and for sub-GeV masses, there is little depth-dependence to $n_\x$. Note that the densities are the same in both panels for $m_\x \gg 1 \ \GeV$, since in this case mCPs are gravitationally bound to the Earth independent of their coupling to terrestrial electromagnetic fields.

We see the enhancing effect of the atmospheric electric field $E_\oplus$ by comparing the right-panel of \Fig{nxCR} to the left-panel. If mCPs directly couple to the SM photon, neglecting $E_\oplus$ is overly pessimistic and we expect a more detailed analysis to result in densities comparable to the right-panel of \Fig{nxCR}. As discussed above, for interactions instead mediated by a dark photon, this local density can be suppressed or enhanced compared to the right-panel of \Fig{nxCR}, but is strictly greater than the densities shown in the left-panel. For instance, the effect of $E_\oplus$ is shorted out if the mCP density grows to the point that it sources a dark electric force of comparable strength, corresponding to a maximal electrically bound mCP density as given above in \Eq{maxCR}. However, even in this case, Earth's magnetic field can still suppress the evaporation of light mCPs. Furthermore, as discussed near \Eqs{darkEoverg}{darkEoverT}, the repulsive effect of the dark electric field can significantly increase the local mCP density near the surface.

It is important to emphasize that the mCP population generated by cosmic rays is irreducible and solely a function of the particle's charge and mass. Hence, testing this local density is equivalent to testing the model itself, akin to an accelerator search. In the rest of this work, we discuss ways to enhance the local density of such mCPs.

\section{Accumulation in the Lab}
\label{sec:compactor}

Here, we explore a simple setup to enhance the local mCP density in the lab relative to the average value on Earth. As depicted in \Fig{cartoon}, this can be achieved by holding an open-air spherical shell of radius $R_0$ at a fixed voltage $\phi_0 < 0$ relative to ground,\footnote{We will often abuse this notation and take $e q_\x \phi_0 > 0$, but it should be understood that $\phi_0$ is an attractive potential.} analogous to a standard Van de Graaff generator.\footnote{Holding the voltage fixed requires continuous charging of the device, since the conductivity of the atmosphere would otherwise discharge the shell on the scale of minutes.} This ``accumulator shell" traps mCPs from the immediate environment and can be independently operated prior to the running of a dedicated experiment to detect the collected mCPs. A detector placed on the interior of the trap is immersed in a larger local mCP density $n_\text{trap}$, enhancing its sensitivity to the terrestrial mCP population. 

Crucially, the shell shields against SM charges, such that only mCPs can penetrate into the interior. In particular, the separation of nuclear and electronic wavefunctions at conducting surfaces generally leads to a potential barrier of order a few eV~\cite{Budker:2021quh}. As a result, room-temperature particles with charge much greater than $300 \ \text{K} / \eV \sim \text{few} \times 10^{-2}$ cannot efficiently penetrate metal barriers. This allows for the effective exclusion of SM ions by the accumulator, while still allowing mCPs to diffuse into its interior. This is the main differentiating factor between mCPs and free SM ions. Furthermore, while mCPs can bind to the macroscopic electrostatic potential of the accumulator, in the parameter space we consider they do not bind with individual SM particles (see, e.g., Ref.~\cite{Pospelov:2020ktu}).\footnote{This same distinguishing behavior between mCPs and SM ions implies that whatever charging mechanism is utilized for the accumulator will not strongly impact the density of mCPs in the interior. For instance, in the case of a Van de Graaff generator, the motorized belt used to carry SM charges onto the shell is unable to form bonds with room-temperature mCPs for $q_\x \ll 0.1$.}

Thus, for $T_\x / (e \phi_0) \ll q_\x \ll T_\x / \eV$, mCPs of temperature $T_\x$ can diffuse into the interior cavity of the charged shell, lose energy via scattering in material, and become electrically bound to its electrostatic potential. In general, terrestrial mCPs accumulate within the trap shell at a rate set by the electric accumulation velocity $V_E$. The density accumulated in a time $t$ is given by
\be
\label{eq:nxtrap}
n_\text{trap} \simeq 3 \, \varepsilon_\text{trap} \, n_\x \, V_E \, t / R_0
~.
\ee
Above, $\varepsilon_\text{trap} \leq 1$ is a dimensionless efficiency factor that is $\order{1}$ for an endless supply of mCPs that are strongly-coupled enough to collisionally-thermalize within the shell (this will be discussed later in \Secs{room}{bound}).
As derived in \App{accumulation}, the accumulation velocity $V_E$ is related to the electric field $E_0 = \phi_0 / R_0$ at the shell's surface by 
\be
\label{eq:VE}
V_E \simeq \frac{e q_\x \, E_0 / m_\x}{\max \big( \Gamma_p^\text{(air)} \, \beta_E \, , \, v_\text{th} / 2 R_0 \big) } \, 
~,
\ee
where $\Gamma_p^\text{(air)}$ is the momentum-exchange rate for mCP-atomic collisions in air, and $v_\text{th} \simeq \sqrt{3 T_\x / m_\x}$ is the mCP thermal velocity at temperature $T_\x$ (which, due to rapid scattering with nuclei, is approximately equal to the environment's temperature). In the denominator of \Eq{VE}, the first term in the ``max" applies to highly-collisional mCPs, whereas the second quantity applies to ballistically free-streaming particles. Thus, the incorporation of the collision rate $\Gamma_p^\text{(air)}$ only serves to reduce the total accumulation rate compared to that in the ballistic regime.

In \Eq{VE}, we have also introduced the dimensionless parameter $\beta_E$, which accounts for screening effects of the shell's electric field from the local environment. As derived in \App{accumulation}, for an electric field that scales as a monopole ($r^{-2}$) or dipole ($r^{-3}$) at far distances, $\beta_E \simeq 1$ or $\beta_E \simeq \sqrt{\pi e q_\x \phi_0 / (2 T_\x)} \gg 1$, respectively. For instance, in the absence of any nearby image charges induced by the shell (such as if the shell is placed indoors near the center of a much larger enclosing conducting structure), the electric field is purely that of a monopole wherever it is nonzero. However, if the shell is placed outdoors near Earth's surface, the conducting nature of the crust generates an image charge below its surface, leading to a dipole-like field at sufficiently far distances. 

As an example, let us consider a monopole-like electric field, corresponding to $\beta_E \simeq 1$. In this case, in the highly-collisional regime, $\Gamma_p^\text{(air)} \gg v_\text{th} / R_0$, \Eq{VE} is simply the terminal velocity of an mCP as it is attracted by the electric field and collides with surrounding air molecules. Instead, for $\Gamma_p^\text{(air)} \ll v_\text{th} / R_0$, this gives $V_E \simeq v_\text{esc}^2 / v_\text{th} \gg v_\text{th}$ where $v_\text{esc}$ is the electrical escape velocity of the shell, corresponding to the Sommerfeld-enhanced capture rate of ballistically free-streaming mCPs in a long-ranged potential.

To calculate the collision rate $\Gamma_p^\text{(air)}$, we follow the approach of Ref.~\cite{Berlin:2023zpn}. In particular, we employ semi-analytic solutions to the Schr\"{o}dinger equation from Refs.~\cite{Tulin:2013teo,Colquhoun:2020adl} with an mCP-atomic interaction governed by the Thomas-Fermi nuclear potential. Parametrically, $\Gamma_p^\text{(air)} \sim (\mu_{\x N} / m_\x) \, n_N^\text{(air)} \, \sigma_T \, v_\text{rel}$, where $n_N^\text{(air)}$ is the number density of ambient atoms in the air, $\sigma_T$ is the transfer cross section for mCP-atomic collisions, and $v_\text{rel}$ is the relative thermal velocity.

In \Eq{nxtrap}, we introduced the efficiency factor $\varepsilon_\text{trap}$. We now further decompose this as
\be
\label{eq:effecgen}
\varepsilon_\text{trap} = \varepsilon_\text{room} \, \varepsilon_\text{bound}
~.
\ee
The first factor $\varepsilon_\text{room}$ accounts for the finite size of the laboratory room (if the accumulator is placed indoors), whereas the second factor $\varepsilon_\text{bound}$ accounts for the ability for mCPs to scatter and become electrically bound to the accumulator. In the following subsections, we discuss each of these in turn. 

\subsection{Indoors vs. Outdoors}
\label{sec:room}

In our estimates, the voltage of the accumulator shell is held fixed for a time $t = 1 \ \yr$. In order to maximize the sensitivity for such timescales, the accumulator should ideally be operated outdoors since the total number of ambient mCPs that it is able to collect is ultimately limited by the size of the largest enclosure dictating the range of the electric field. 

For instance, let us take the electric field of the accumulator to be confined to a room of radius $R_\text{room}$, whose walls are of thickness $\Delta R_\text{room}$, and with a ventilation opening of area $A_\text{vent}$ that pushes air (and collisionally-coupled mCPs) into the room at velocity $V_\text{vent}$. The total number $N_\text{trap}$ of mCPs collected in the accumulator over a time $t$ is limited by the  number of particles originating in the room, entering through the ventilation, and diffusing in through the walls.  As derived in \App{diffusion}, this requires imposing that
\begin{align}
\label{eq:room}
N_\text{trap} &\lesssim n_\x \, 4 \pi R_\text{room}^2 \, t \,  \bigg( \frac{R_\text{room}}{3 t} \,  + \frac{A_\text{vent} \, V_\text{vent}}{4 \pi R_\text{room}^2} 
\nl
& +   \frac{D_\x^\text{(wall)}}{\Delta R_\text{room}} + \frac{V_g^{(\text{air})}}{4} \bigg)
~,
\end{align}
where $D_\x^\text{(wall)} = T_\x /  (m_\x \, \Gamma_p^{\text{(wall)}})$ is the mCP diffusion coefficient in the room's walls and $V_g^{(\text{air})} = g_\oplus / \Gamma_p^{(\text{air})}$ is the mCP's drift velocity in Earth's gravitational field $g_\oplus$.\footnote{Note that $V_g^{(\text{air})}$ is the gravitational drift velocity in air, not the wall. This accounts for the fact that in steady-state, conservation of flux implies that mCPs can develop an overdensity in the ceiling of the building at the level of $V_g^{(\text{air})} / V_g^{(\text{wall})} \gg 1$. This occurs provided that mCPs pass through the ceiling before diffusing transversely to the building's side-walls, which is possible if the transverse length scale of the roof is greater than $\sim \sqrt{\Delta R_\text{room} \, v_\text{th} \, \Gamma_p^{(\text{wall})}} / \Gamma_p^{(\text{air})}$. We have checked that for reasonable parameters, this is satisfied throughout the majority of the relevant parameter space.  We do not incorporate the loss of mCPs from the sides of the roof in our estimates, but it would have at most a minor effect and only for the largest masses and smallest charges that we consider.} 
We incorporate the requirement of \Eq{room} into the efficiency factor  of \Eq{effecgen} by taking
\begin{align}
\label{eq:epsroom}
\varepsilon_\text{room} = \min \big[ \, 1 ~ , ~  V_\text{room}^\text{eff} / V_E \, \big]
~,
\end{align}
where we defined the effective velocity
\begin{align}
\label{eq:Vroom}
V_\text{room}^\text{eff} &\equiv \bigg( \frac{R_\text{room}}{R_0} \bigg)^2 \, \bigg( \frac{R_\text{room}}{3 t} + \frac{A_\text{vent} \, V_\text{vent}}{4 \pi R_\text{room}^2}
\nl
&+ \frac{D_\x^{(\text{wall})}}{\Delta R_\text{room}} + \frac{V_g^{(\text{air})}}{4}  \bigg)
~.
\end{align}

It is worth examining the various terms of \Eq{Vroom} for a concrete setup consisting of an accumulator operated for a time $t = 1 \ \yr$ in a room of radius $R_\text{room} = 10 \ \m$, walls of thickness $\Delta R_\text{room} = 1 \ \m$, and a vent of area $A_\text{area} = 1 \ \m^2$ and air speed $V_\text{vent} = 10 \ \m / \text{s}$. In this case, the first term in \Eq{Vroom} is negligible compared to the second term,  $R_\text{room} / t \sim 10^{-5} \ \cm / \text{sec} \ll A_\text{vent} \, V_\text{vent} / (4 \pi R_\text{room}^2) \sim 1 \ \cm / \text{sec}$. For sufficiently small charges, the third and fourth terms of \Eq{Vroom} take simple approximate forms. In particular, for $q_\x \ll 10^{-4}$, we find that for $m_\x \ll 1 \ \GeV$, 
\be
\frac{D_\x^{(\text{wall})}}{\Delta R_\text{room}} \sim 10^8 \  \frac{\cm}{\text{s}} \times \bigg( \frac{10^{-6}}{q_\x} \bigg)^2 \, \bigg( \frac{\MeV}{m_\x}\bigg)^{5/2}
\ee
and
\be
\frac{V_g^{(\text{air})}}{4} \sim 10^2 \  \frac{\cm}{\text{s}} \times \bigg( \frac{10^{-6}}{q_\x} \bigg)^2 \, \bigg( \frac{\MeV}{m_\x}\bigg)^{3/2}
~,
\ee
while for $m_\x \gg 1 \ \GeV$ we have
\be
\frac{D_\x^{(\text{wall})}}{\Delta R_\text{room}} \sim 10^{-1} \  \frac{\cm}{\text{s}} \times \bigg( \frac{10^{-6}}{q_\x} \bigg)^2 
\ee
and
\be
\frac{V_g^{(\text{air})}}{4} \sim 10^{-1} \  \frac{\cm}{\text{s}} \times \bigg( \frac{10^{-6}}{q_\x} \bigg)^2 \, \bigg( \frac{m_\x}{10^2 \ \GeV}\bigg)
~.
\ee

In the case that diffusion through the walls, corresponding to the third term of \Eq{Vroom}, dominates over the other terms, it is a limiting experimental factor if the naive rate of electrical accumulation in the trap is larger than the rate of new particles diffusing into the room, i.e., $V_E \, R_0^2 \gtrsim D_\x^{(\text{wall})} \, R_\text{room}^2 / \Delta R_\text{room}$. This occurs for a room smaller than
\begin{align}
R_\text{room} &\lesssim \sqrt{R_0 \, \Delta R_\text{room} \, n_N^\text{(wall)} / n_N^\text{(air)}} ~ v_\text{esc} / v_\text{th}
\nl
&\simeq 10 \ \km \times  \sqrt{\frac{q_\x}{10^{-3}} \, \frac{\phi_0}{\text{MV}} \, \frac{R_0}{\m}\, \frac{\Delta R_\text{room}}{\m}}
~,
\end{align}
where $n_N^\text{(wall)}$ and $n_N^\text{(air)}$ are the nuclear density of material in the wall or air, respectively. Thus, operating the trap indoors degrades the accumulated density for any reasonably sized indoor environment. Note, however, that this can be mitigated by efficient air ventilation (especially for larger charges) or running the accumulator outdoors well before operation of an enclosed mCP experiment.

\subsection{Electrical Trapping}
\label{sec:bound}

In order for an mCP to become electrically bound to the accumulator, we will take the condition that the energy $e q_\x \phi_0 \gg T_\x$ gained in the shell's potential well must be lost through collisions with atoms nearby the shell. As discussed in Ref.~\cite{Berlin:2023zpn}, the energy-exchange rate $\Gamma_E$ for mCP-atomic scattering is related to the momentum-exchange rate $\Gamma_p$  by $\Gamma_E \simeq (\mu_{\x N} / m_N) \, \Gamma_p$, where $m_N$ is the atomic mass. An mCP can lose an $\order{1}$ fraction of its energy in a region of material of length $\Delta R$ if
\be
\label{eq:therm1}
\Gamma_E \, \Delta R^2 / D_\x \gtrsim 1
~.
\ee
For example, we find that in a meter of room-temperature iron, \Eq{therm1} requires $q_\x \gtrsim 10^{-5} \times (\MeV / m_\x)^{5/4}$ for $m_\x \ll 1 \ \GeV$ and $q_\x \gtrsim  10^{-9} \times (m_\x / 10^2 \ \GeV)^{1/4}$ for $m_\x \gg 1 \ \GeV$.  

The probability $\text{Pr}_\text{th}^{(\text{shell})}$ that mCPs thermalize in the shell can be $\order{1}$ if \Eq{therm1} is satisfied after summing over all elements of the accumulator. In this case, the specific form of $\text{Pr}_\text{th}^\text{(shell)}$ then depends on the mCP's collisional mean free path in air, $\ell_p^{(\text{air})} \simeq v_\text{th} / \Gamma_p^{(\text{air})}$. For instance, if $\ell_p^{(\text{air})} \lesssim R_0$, mCPs undergo a tightly-coupled random walk with a strong bias to remain close to the shell, such that $\text{Pr}_\text{th}^{(\text{shell})} \simeq 1$ if \Eq{therm1} is satisfied. 

Instead, if $\ell_p^{(\text{air})} \gg R_0$, then even mCPs that encounter the shell may reflect off and free-stream away a significant distance before fully thermalizing. In this case, such mCPs may escape entirely by diffusing out of the enclosing room or being carried away by a gust of wind. This is increasingly likely  for lower mass particles, since many momentum-exchanges ($\sim m_N / \mu_{\x N}$) are required for efficient energy transfer. As a result, the chance that mCPs reflect off of the accumulator before exchanging a meaninful amount of energy becomes significant for $m_\x \ll m_N$. As derived in Ref.~\cite{Berlin:2023zpn}, we can account for this by taking $\text{Pr}_\text{th}^{(\text{shell})} \simeq \text{erf}\big( \sqrt{3 m_\x / 4 m_N} \big)$ if \Eq{therm1} is satisfied and $\ell_p^{(\text{air})} \gg R_0$.

Along these lines, also note that an mCP only needs to lose a small fraction of its incoming energy to become electrically bound, which is weaker than the $\order{1}$ energy-loss requirement of \Eq{therm1}. However, thermalized mCPs satisfying \Eq{therm1} remain close to the accumulator, and are less subject to effects of the environment at further distances that could remove them from the immediate vicinity. In this sense, our analysis is conservative and ignores mCPs that become weakly bound to the accumulator and are able to propagate out to far distances. 

Even after mCPs efficiently thermalize with the shell, they may still evaporate. After a time $t$, the fraction of mCPs that remain bound to the shell can be approximated by $(1 - e^{-\Gamma_\text{evap} \, t}) / (\Gamma_\text{evap} \, t)$, where the evaporation rate is roughly~\cite{Berlin:2023zpn}
\be
\label{eq:evap}
\Gamma_\text{evap} \simeq \frac{3}{2 R_0} \, \sqrt{\frac{2 T_\x}{\pi m_\x}} \, \bigg( 1 + \frac{e q_\x \, \phi_0}{T_\x} \bigg) \, e^{- e q_\x \phi_0 / T_\x}
~.
\ee
\begin{figure*}[t!]
\centering
\includegraphics[width=0.495 \textwidth]{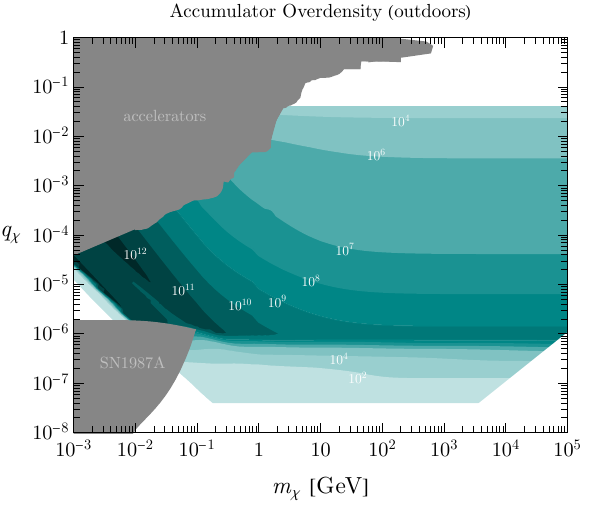}
\includegraphics[width=0.495 \textwidth]{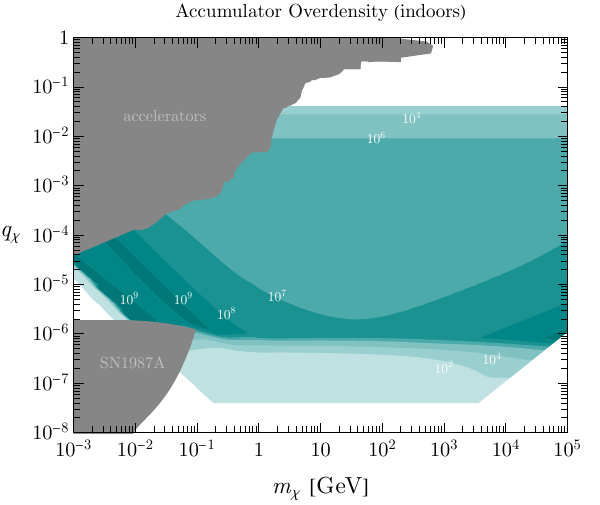}
\caption{The overdensity $n_\text{trap} / n_\x$  of millicharged particles accumulated by an electric trap (compared to the average nearby terrestrial density), as a function of the millicharge $q_\x$ and mass $m_\x$. The trap consists of a room-temperature shell of radius $R_0 = 2 \ \m$, held at a voltage of $\phi_0 = - 1 \ \MV$ for a time $t = 1 \ \yr$. We consider a setup operated outdoors (left) or indoors (right). In gray, we also show existing limits from accelerator probes~\cite{Davidson:2000hf,Haas:2014dda,Prinz:1998ua,ArgoNeuT:2019ckq,milliQan:2021lne,ArguellesDelgado:2021lek,PBC:2025sny,CMS:2024eyx,Alcott:2025rxn} and SN1987A~\cite{Chang:2018rso}. The region of parameter space for which overdensities can develop is limited to a finite region of masses and couplings: at small masses, millicharged particles are less efficient at thermalizing in the shell through scattering, and at large masses Earth's gravitational field overtakes the effect of the accumulator's electric field; for small charges, millicharged particles are not electrostatically bound to the shell, whereas for large charges they form bound states with electrons and cannot freely penetrate conducting surfaces.}
\label{fig:trap}
\end{figure*}

Thus far, we have taken the electric force from the accumulator to be the dominant long-ranged force acting on the terrestrial mCP population. This is valid provided that the electric force from the shell dominates over Earth's gravitational field, i.e.,
\be
\label{eq:grav1}
q_\x \gtrsim \frac{m_\x \, g_\oplus}{E_0} \sim 10^{-12} \times \bigg(\frac{m_\x}{\GeV} \bigg) \bigg(\frac{\kV / \cm}{E_0} \bigg)
~.
\ee
If this is not satisfied, mCPs are gravitationally stripped from the shell (which could be overcome by running the setup in free fall). In our analysis, we refrain from considering couplings less than three times this critical value. From \Eq{grav1}, we see that this is unimportant for light mCPs produced by cosmic rays but is potentially relevant for mCP DM with $m_\x \gtrsim 1 \ \TeV$.

Even for mCPs that are not stripped from the shell, Earth's gravitational field can still significantly perturb the distribution of mCPs in the interior. For instance, instead of being uniformly distributed throughout the shell, mCPs will tend to sit towards the bottom of the interior cavity if $m_\x \gtrsim T_\x / (g_\oplus \, R_0) \sim 250 \ \TeV \times (T_\x / 300 \ \text{K}) \, (\m / R_0)$. For masses a few orders of magnitude larger, mCPs reside within the barrier of the shell, making their detection increasingly difficult. Since \Eq{nxtrap} assumes a uniformly distributed charged density, we refrain from considering mCPs heavier than $100 \ \TeV$. 

We incorporate the requirements of \Eqss{therm1}{evap}{grav1} 
into the efficiency factor of \Eq{effecgen} by taking
\be
\label{eq:epsbound}
\varepsilon_\text{bound} =  \text{Pr}_\text{th}^{(\text{shell})} ~ \frac{1 - e^{-\Gamma_\text{evap} \, t}}{\Gamma_\text{evap} \, t} 
~,
\ee
if $q_\x > T_\x / (e \phi_0)$ and \Eqs{therm1}{grav1} are satisfied, and $\varepsilon_\text{bound} = 0$ if not.

Throughout this discussion, we have treated mCPs as an ideal gas. However, this may not always be the case. Positively-charged mCPs, which remain unbound with nuclei, move more freely than negatively-charged mCPs. Thus, the former accumulate first inside the shell. As the density of positively-charged mCPs increases, their electrical repulsion pushes them outwards towards the shell's interior boundary. This effect is negligible when the self-potential is less than the mCP's kinetic energy, such that the density remains approximately uniformly distributed throughout the interior cavity. However, if enough positively-charged mCPs accumulate, the energy gained by another mCP occupying the central region will eventually exceed the temperature.  

Therefore, if only positively-charged mCPs accumulate, this can hinder the growth of the millicharge density near a detector placed in the center of the cavity, since, in this case, all further accumulated particles will tend to occupy a small region near the shell wall. However, the accumulating population of positively-charged mCPs will also attract negatively-charged SM particles (electrons or ions). This equilibrium configuration corresponds to zero net charge density in the interior cavity, with the positive millicharge being canceled by negative SM charge. On timescales much greater than a day, we expect that this equilibrium configuration can be attained, since all materials composing the apparatus  have some finite conductivity. Of course, if needed, it should also be possible to purposely engineer a faster way for these negative SM charges to reach the interior. This neutralizing SM population would generally not interfere with the mCP detectors discussed in \Sec{detection} and Ref.~\cite{forthcoming} (such as ion traps or Cavendish experiments) because of their differing ability to penetrate dense material.

\section{Results}
\label{sec:results}

Using the formalism described above, we determine the mCP overdensity $n_\text{trap} / n_\x$ collected by the accumulator over a time $t = 1 \ \yr$. In particular, we consider a room-temperature setup consisting of a trap shell of radius $R_0 = 2 \ \m$ and thickness $1 \ \mm$ that encloses a solid sphere of radius $0.5 \ \m$, the latter of which enhances the likelihood for mCPs to collisionally thermalize and become electrically bound (see \Eq{therm1}). For concreteness, we take all material to be made of iron. The outer shell is held at a fixed voltage of $\phi_0 = - 1 \ \text{MV}$ relative to ground (the sign of $\phi_0 < 0$ is chosen to target positively-charged mCPs, which remain unbound with atomic nuclei, as discussed in \Sec{overdensity}). Note that these parameters are well within what is achievable today, comparable to large existing open-air Van de Graaff generators on public display~\cite{bostonmuseum}.

In \Fig{trap}, we show the overdensity $n_\text{trap} / n_\x$ of mCPs accumulated for different experimental configurations, which shows that this setup can enhance the local density by as much as twelve orders of magnitude. We take the accumulator to be placed outdoors (left) or indoors (right); in the latter case, we take the enclosing room to be of radius $R_\text{room} = 10 \ \m$ with walls of thickness $\Delta R_\text{room} = 30 \ \cm$ and with a vent of area $A_\text{vent} = 1 \ \m^2$ and air speed $V_\text{vent} = 10 \ \m / \text{s}$. 

For either of these setups, the overdensity is suppressed at very small or very large masses. For $m_\x \ll 10 \ \GeV$, the kinematic mismatch between the mCP mass and the mass of typical nuclei means that mCPs are less able to efficiently exchange energy with material in the interior of the accumulator (see \Eq{therm1}). Instead, the dominant suppression at high masses arises from the fact that Earth's gravitational field overtakes the effect of the trap's electric field, hindering the collection of mCPs (see \Eq{grav1}).

The overdensities are also suppressed at sufficiently small or large charges. For very small charges, the evaporation rate $\Gamma_\text{evap}$ of \Eq{evap} becomes non-negligible and mCPs do not efficiently bind to the shell. This can be seen in in \Fig{trap}, which shows a falloff in $n_\text{trap} / n_\x$ for charges only slightly greater than the critical value $q_\x \sim T_\x / (e \phi_0)$. 

Instead, for very large charges, the overdensity is ultimately limited by the formation of mCP-electron bound states (see \Sec{overdensity}) as well as the inability of positively-charged mCPs to penetrate the repulsive voltage difference present near the surface of any conducting material~\cite{Budker:2021quh} (see the discussion above \Eq{nxtrap}). For the former, we refrain from considering couplings larger than $\sim \sqrt{T_\x / \text{Ry}}$, while for the latter, we incorporate the fraction of mCPs that are able to overtake a surface-barrier of $\sim 5 \ \eV$ as a simple Boltzmann factor dictated by the mCP temperature.

As discussed in \Sec{compactor}, if the accumulator is placed outdoors, then image charges generated by Earth's crust results in the accumulator's electric field scaling as a dipole at far distances (this enters in the form for $\beta_E$ in \Eq{VE}). Despite this suppression, an outdoor setup accumulates a larger overdensity due to the approximately endless supply of mCPs nearby. This is evident in  \Fig{trap}, which shows that the accumulated overdensity is suppressed by a couple orders of magnitude for a setup operated indoors. Note that for large masses, the gravitational pull of mCPs through the ceiling of the enclosure mitigates this suppression. 

\subsection{Experimental Variations}
\label{sec:variations}

Here, we briefly comment on experimental variations away from the baseline parameters adopted above. For instance, although we have focused on a geometrically large accumulator, parametrically smaller setups retain many of the benefits of this proposal. In fact, for an electric field $E_0$ fixed to a value below the dielectric strength of air ($\sim 3 \ \MV / \m$), the scaling $n_\text{trap} \propto E_0 / R_0$ in \Eq{nxtrap} implies that smaller radii $R_0$ can even enhance the accumulated mCP density. However, in this case $\phi_0$ is also reduced, increasing the minimum charge for which mCPs can efficiently bind to the accumulator. Although a smaller accumulator possesses less material to facilitate mCP energy-loss through scattering, only a small amount of material is needed for the larger values of $q_\x$ considered in this work. As a result, a compact table-top version of our proposal involves limited drawbacks and is ideal for a pilot experiment.  

The accumulator could also be run at cryogenic temperatures, which offers various advantages and disadvantages. At lower temperatures, more feebly-coupled mCPs are trapped at a fixed voltage, enhancing the accumulated density at small coupling. However, the introduction of a specialized refrigeration system would require additional shielding, likely limiting the range of the accumulator's electric field. We therefore do not consider this possibility further.

In the analysis above, we showed that the accumulation of mCPs can be hindered if operated indoors, such as if the accumulator is enclosed within a larger conducting structure, since this limits the range of the shell's electric field. Indeed, this is what was assumed for our setup labeled ``indoors," which showed reduced accumulation overdensities in \Fig{trap}.  However, this is not necessarily the case for a realistic experiment. For instance, this suppression can be significantly mitigated if the walls of the surrounding enclosure are partially insulating, such that the accumulator's electric field can penetrate out to substantially larger distances. Thus, if  an $\order{1}$ solid-angle fraction of the enclosure involves insulating material, an indoor setup will instead accumulate overdensities at a level more comparable to what is shown for the ``outdoors" projection in \Fig{trap}.

\begin{figure*}[t]
\centering
\includegraphics[width=0.495 \textwidth]{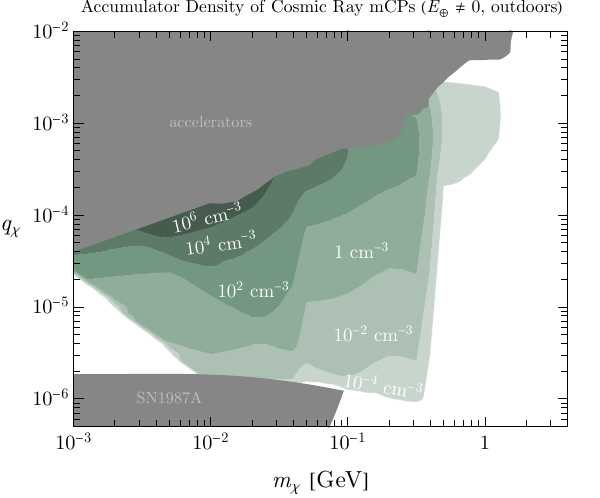}
\includegraphics[width=0.495 \textwidth]{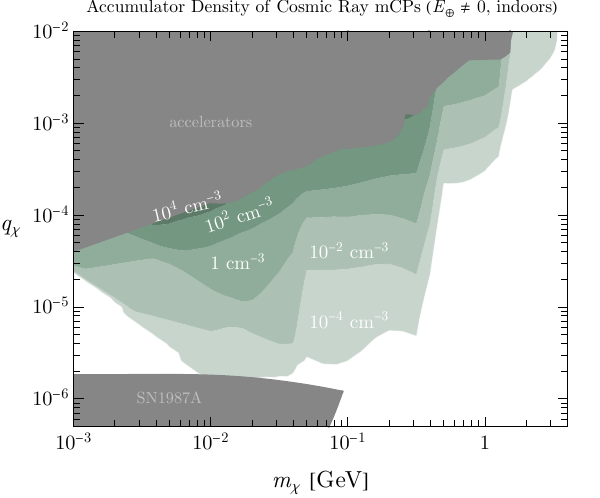}
\includegraphics[width=0.495 \textwidth]{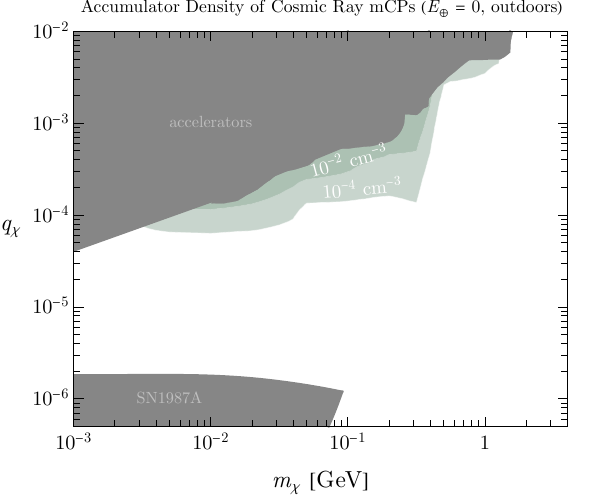}
\includegraphics[width=0.495 \textwidth]{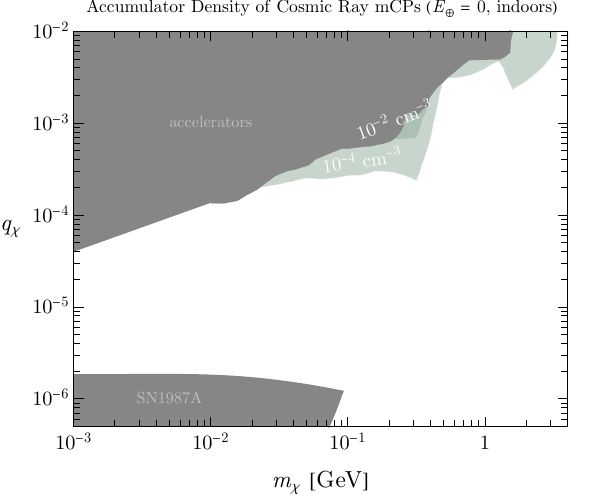}
\caption{Using the results of \Figs{nxCR}{trap}, we determine the accumulated density $n_\text{trap}$ of positively-charged millicharged particles initially produced by cosmic rays. In the top row, the notation ``$E_\oplus \neq 0$" indicates that we take the millicharges to efficiently couple to the atmospheric electric field, whereas in the bottom row ``$E_\oplus = 0$" indicates that the role of the atmospheric electric field is neglected (this is shown for the purpose of illustrating the enhancing effect of $E_\oplus$). For millicharged particles that couple directly to the photon, we expect that detailed modeling will result in densities similar to the results for ``$E_\oplus \neq 0$." These results can be significantly modified if the interaction is mediated by a kinetically-mixed dark photon, potentially yielding densities that are parametrically larger than those shown here, depending on the specific model parameters. In gray, we also show existing limits from accelerator probes~\cite{Davidson:2000hf,Haas:2014dda,Prinz:1998ua,ArgoNeuT:2019ckq,milliQan:2021lne,ArguellesDelgado:2021lek,PBC:2025sny,CMS:2024eyx,Alcott:2025rxn} and SN1987A~\cite{Chang:2018rso}. In the left-column, ``outdoors" means that the accumulator is operated $ 1 \ \text{m}$ above the crust, whereas in the right-column ``indoors" means that the accumulator is operated inside of a room $1 \ \text{m}$ below the crust (this latter scenario is qualitatively similar to operating inside of a conducting building slightly above ground).}
\label{fig:TrapCRnx}
\end{figure*}
%

\subsection{Dark Matter and Cosmic Ray Millicharges}

\Fig{trap} describes the overdensity of mCPs that an electrified shell can accumulate compared to the average terrestrial density $n_\x$, regardless of the origin of $n_\x$. As mentioned in \Secs{mCPDM}{CRpop}, this ambient terrestrial population can arise from, e.g., a Galactic DM subcomponent, or locally from cosmic ray air showers. In the former case, Ref.~\cite{Berlin:2023zpn} showed that $n_\x$ can be larger than the Galactic density of mCPs by as much as $\sim 10^{16}$ over a wide-range of couplings and masses. The results shown in \Fig{trap} therefore imply that, compared to the Galactic value, the density of an mCP DM subcomponent can be locally enhanced in the laboratory by as much as 28 orders of magnitude.

The irreducible terrestrial density of mCPs produced by cosmic rays was shown in \Fig{nxCR}, depending on whether or not mCPs efficiently couple to the atmospheric electric field. These results can be multiplied by the overdensity in \Fig{trap} to obtain the irreducible mCP density accumulated within an electrified shell.  This is shown in \Fig{TrapCRnx} for an outdoor or indoor setup. In either case, the accumulated density of mCPs can be as large as $n_\x \sim 10^6 \ \cm^{-3}$ and $\sim 10^4 \ \cm^{-3}$, respectively, if mCPs efficiently couple to the atmospheric electric field, and $n_\x \sim 10^{-2} \ \cm^{-3}$ if not. 

For the outdoor setup, we take the accumulator to be placed at a height $h \sim 1 \ \m$ above the ground. However, a population of mCPs coupled to the atmospheric electric field resides dominantly below the crust, extending a distance $\sim T_\x / (e q_\x \, E_\oplus)$ above the surface. We conservatively incorporate this effect as a Boltzmann factor $n_\x \propto \text{exp}[- e q_\x E_\oplus h / T_\x]$, which suppresses $n_\text{trap}$ in the top-left panel of \Fig{TrapCRnx} for $q_\x \gtrsim \text{few} \times 10^{-4}$. Note, however, that this is overly pessimistic, as the value of the electric field at the crust is dominated by $E_0 \gg E_\oplus$ near the accumulator. Instead, for the indoor setup, we assume that the accumulator is placed $1 \ \m$ underground, such that there is no suppression for such couplings (this is qualitatively similar to operating inside of a conducting building slightly above ground).

\subsection{Kinetically-Mixed Models} 
For most of the above discussion, we treated mCPs as genuinely charged under the SM photon. However, when the  interaction is mediated by an ultralight dark photon, such that $q_\x = \eps \, e^\p / e$, mCPs effectively couple to both the dark and visible photons. In \Sec{overdensity}, we discussed how such additional interactions can modify the population of mCPs trapped on Earth by the atmospheric electric field. For instance, the repulsive effect of the dark electric field can result in mCPs residing predominantly near Earth's surface, parametrically increasing the local density compared to what was assumed in generating \Fig{TrapCRnx}. Similar considerations apply to the role of the accumulator as well. In this case, the accumulated mCPs backreact and prevent further accumulation of charge once they source a \emph{dark} electric field $E^\p$ whose force on the incoming mCPs overcomes that of the visible field. This occurs once 
\be
e^\p \, E^\p \sim e^{\p \, 2}  \, n_\text{trap} \, R_0 \sim e q_\x \, E_0
~,
\ee
corresponding to an mCP-induced visible electric field 
\be
E_\x \sim e q_\x \, n_\text{trap}  \, R_0 \sim (e q_\x / e^\p)^2 \, E_0 \sim \eps^2 \, E_0
~.
\ee
As a result, accumulation of charge quenches once the mCPs screen an $\eps^2$ fraction of the shell's visible electric field. For accumulation of a single charge species, analogous to \Eq{maxCR}, this occurs when the accumulated mCP density grows to
\be
\label{eq:backreact}
n_\text{trap} \sim \frac{3 e q_\x \, \phi_0}{e^{\p \, 2} \, R_0^2} 
\sim 10^6 \ \cm^{-3} \times \bigg( \frac{q_\x}{10^{-3}} \bigg) \, \bigg( \frac{10^{-1}}{e^\p} \bigg)^2
~,
\ee
for $R_0 = 2 \ \m$ and $\phi_0 = 1 \ \MV$. Thus, at the perturbative limit $\eps \lesssim 1$ where $e^\p \sim e q_\x$, this model-dependence does not affect the results of \Fig{TrapCRnx}. However, for large values of the dark gauge coupling, the accumulated density could saturate before reaching the largest values of $n_\text{trap}$ shown in \Fig{TrapCRnx}. 

Even if the effect of $E_0$ overcomes that of $E^\prime$ and mCPs are successfully attracted to the accumulator, the force from $E^\prime$ may repel the mCPs to the shell's interior boundary. This depends on the mCP's kinetic energy and occurs for densities larger than
\be
\label{eq:pushtoedge}
n_\text{trap} \sim \frac{T_\x}{(e^\p \, R_0)^2} \sim 1 \ \cm^{-3} \times \bigg( \frac{0.1}{e^\p} \bigg)^2
~,
\ee
for $R_0 = 2 \ \m$ and $T_\x = 300 \ \text{K}$. This corresponds to the largest density that freely diffuses through the interior cavity of the shell. 
Note, though, that in the next section (as well as in Ref.~\cite{forthcoming}) we discuss detection strategies that are sensitive to densities much smaller than those in \Eqs{backreact}{pushtoedge}, so this model-dependence does not strongly impact the projected experimental sensitivity.\footnote{For densities smaller than \Eq{pushtoedge}, two-stream plasma instabilities do not develop since the length scale $\sim m_D^{\p \, -1}$ of the fastest growing mode is larger than $R_0$, where $m_D^\p$ is the mCP's contribution to the dark photon's Debye mass~\cite{Lasenby:2020rlf,DeRocco:2024ifs}.}

As discussed in \Sec{bound}, the accumulation of positively-charged mCPs inside the shell will likely bring in excess negatively-charged SM particles (e.g., electrons) to cancel the net visible charge.  However, in a kinetically-mixed model the situation is more complicated.  An overdensity of positive millicharges not only has a net visible charge but also a net dark  charge that is generically much greater, since $e^\p \gg \eps \, e^\p = e q_\x$. These positive mCPs thus may also attract negative mCPs that cancel the total dark charge.  Although the negative mCPs may be bound to nuclei (which is possible in certain regions of parameter space), they may still diffuse into the interior of the shell (e.g., if the vacuum is imperfect), and build up to a density that cancels the dark field in the interior cavity of the shell.

Therefore, we should actually regard \Eqs{backreact}{pushtoedge} as being limits on the net mCP \emph{asymmetry} rather than on the total number density; a symmetric millicharge population can accumulate in the interior cavity to a total number density that is larger than \Eqs{backreact}{pushtoedge}, since as negative mCPs enter, they reduce the total dark charge, allowing more positive mCPs to enter as well. This is relevant since many detector technologies (e.g., ion traps or Cavendish experiments, as discussed below) are sensitive to the \emph{total} number density of mCPs, not to the asymmetry.  Thus, the signals in such detectors could be much larger than the naive limits from \Eqs{backreact}{pushtoedge} (e.g., if the ambient density of millicharges exceeds those limits). However, note that this larger density is irrelevant if the detectors are already sensitive to densities below those in \Eqs{backreact}{pushtoedge}, which we show is possible in Ref.~\cite{forthcoming}. Regardless, for detectors that are not sensitive to the densities in \Eqs{backreact}{pushtoedge}, it is advantageous to pursue schemes that facilitate both species of millicharge to enter the accumulator, since this will significantly boost their ultimate discovery potential. 

Aside from the collective effects discussed above, momentum-exchange from scattering between single pairs of $\x^+$ and $\x^-$ introduces an additional drag on the accumulating positively-charged mCPs. This amounts to including an additional contribution to the momentum-exchange rate $\Gamma_p$ from such processes, and so is only important when this dominates over mCP-SM scattering. Since the local mCP number density is typically parametrically smaller than the ambient density of atoms, this is a small effect for most scenarios considered here. More generally, mCP self-scattering is irrelevant when the corresponding mean free path is greater than the size of the experiment, $(\alpha^{\p \, 2} n_\x / T_\x^2) \, R_0 \lesssim 1$, which is a weaker requirement than demanding $n_\x$ to be smaller than the density shown in \Eq{pushtoedge} by a factor of $T_\x \, R_0 \sim 10^5$.

\subsection{Detection Prospects} 
\label{sec:detection}

Several detection strategies can be employed to search for the mCP population accumulated within the electrostatic trap. Here, we provide a short overview of various approaches. More details regarding one of these strategies is provided in a companion paper~\cite{forthcoming}, while other approaches will be developed further in upcoming work. 

A well-established method for measuring ambient mCPs involves ion traps~\cite{Carney:2021irt,Budker:2021quh}. These experiments exploit the stability of trapped ions in shallow potential wells to search for heating from mCP Rutherford scattering. Ion traps have been shown to be sensitive to mCP densities as small as $n_\x \sim 1 \ \cm^{-3}$ for GeV-scale masses~\cite{Budker:2021quh}. The number densities shown in \Fig{TrapCRnx} fall within this existing sensitivity range, provided such an ion trap could be implemented inside the electrostatic accumulator. We defer a detailed assessment of an ion-trap–based detector housed within an accumulator to future work.

A second method for detecting mCPs relies on their in-medium effects on the photon's dispersion relation. This principle has previously been applied in the context of resonant detectors to search for mCP DM~\cite{Berlin:2019uco,Berlin:2023gvx} and for mCPs produced in the Sun~\cite{Berlin:2021kcm,Berlin:2024ewa}. In a companion paper~\cite{forthcoming}, we investigate how a local thermalized population of mCPs can be detected using Cavendish-type experiments originally designed to probe deviations from Gauss's law. In particular, century-old experiments measuring electric fields inside of a shell driven with an oscillating voltage achieved sensitivity to electric fields that could be sourced by charge densities as small as $q_\x \, n_\x \sim 10^{-4} \ \cm^{-3}$~\cite{Plimpton:1936ont}. Reproducing such experiments within an electrostatic accumulator would thus gain orders of magnitude of new sensitivity to terrestrial mCPs, including those produced by cosmic rays. 

A third method for detecting a thermalized mCP population was proposed in Ref.~\cite{Pospelov:2020ktu}. In this case, the potential difference of a DC electrostatic accelerator is used to accelerate thermalized mCPs to higher energies. These accelerated mCPs are then directed into existing low-threshold sensors. As shown in Ref.~\cite{Pospelov:2020ktu}, this approach is sensitive to cosmic-ray–produced mCPs even without the use of an electrostatic trap. The feasibility of integrating such a DC accelerator within the trap considered here, to further enhance its sensitivity, will be investigated in future work.

\section{Conclusion}
\label{sec:conclusion}

In this work, we have demonstrated that a conducting shell held at a large voltage (such as a Van de Graaff generator) can act as a powerful accumulator for a local room-temperature population of millicharged particles (mCPs). Such a device is capable of enhancing the local density of mCPs by up to twelve orders of magnitude compared to the ambient terrestrial population. This dramatic amplification should enable a broad range of existing detection strategies—such as ion traps, Cavendish-type experiments, and electrostatic acceleration techniques—to achieve sensitivity well beyond current limits.

The overdensity generated within the accumulator extends to both cosmologically-produced mCPs and the population locally sourced by cosmic rays. The latter represents an irreducible population (i.e., probing it is a direct probe of the particle, agnostic to its cosmological abundance). For realistic experimental configurations, we find that accumulated densities can fall within the reach of precision ion-trap measurements, Cavendish experiments, and electrostatic accelerator experiments, highlighting the near-term feasibility of probing previously unexplored parameter space.

Our results open a new direction in the search for light particles beyond the Standard Model and is a complementary approach to accelerator-based searches. In a companion paper~\cite{forthcoming}, we show how a simple Cavendish test housed within an accumulator can probe currently unexplored parameter space, outperforming future accelerator searches for sub-GeV mCPs. Future work will focus on the integration of other concrete detection strategies within the accumulator and potential optimizations such as cryogenic operation. Together, these efforts chart a clear experimental path toward the discovery or exclusion of mCPs across a vast range of masses and couplings.


\section*{Acknowledgements}

We thank Paddy Fox and Kevin Kelly for valuable conversations. This manuscript has been authored in part by Fermi Forward Discovery Group, LLC under Contract No. 89243024CSC000002 with the U.S. Department of Energy, Office of Science, Office of High Energy Physics. This material is based upon work supported by the U.S. Department of Energy, Office of Science, National Quantum Information Science Research Centers, Superconducting Quantum Materials and Systems Center (SQMS) under contract number DE-AC02-07CH11359. 
This work was supported in part by NSF Grant No.~PHY-2310429, Simons Investigator Award No.~824870,  the Gordon and Betty Moore Foundation Grant No.~GBMF7946, and the John Templeton Foundation Award No.~63595.

\bibliography{main}


\clearpage
\begin{appendices}
\onecolumngrid

\section{Accumulation}
\label{app:accumulation}

In this appendix, we derive the form of the accumulation velocity $V_E$ in \Eq{VE}. Let us first investigate the highly-collisional regime, in which case mCPs can be approximated as an ideal fluid. The Euler equation in the presence of external electric $\Ev$ and gravitational $\gvec$ fields is then
\be
\label{eq:Euler1}
\dt \Vv_\x + (\Vv_\x \cdot \grad) \, \Vv_\x + \Gamma_p \, \Vv_\x = \frac{e q_\x}{m_\x} \, \Ev + \gvec - \frac{\grad P_\x}{m_\x \, n_\x}
~,
\ee
where $\Vv_\x$ is the mCP bulk velocity and $P_\x \simeq T_\x \, n_\x$ is the mCP pressure. As derived in Ref.~\cite{Berlin:2023zpn}, the momentum-exchange rate from mCP-atomic collisions is
\be
\Gamma_p \simeq n_N \times
\begin{cases}
3 \, \bar{v}_0^2 \, \langle v_\text{rel} / \sigma_T \rangle^{-1} & (m_\x \ll m_N)
\\
(m_N / m_\x) \, (3 \bar{v}_0^2)^{-1} \, \langle \sigma_T v_\text{rel}^3 \rangle  & (m_\x \gg m_N)
~,
\end{cases}
\ee
where $\sigma_T$ is the transfer cross section, the brackets refer to a thermal average of the relative velocity $v_\text{rel}$ with a Maxwellian distribution of variance $\bar{v}_0^2 \equiv T_\x / m_\x + T_N / m_N$, and $n_N$, $m_N$ and $T_N$ are the nuclear density, mass, and temperature, respectively. In the highly-collisional limit, we can ignore the first two terms on the left-hand side of \Eq{Euler1}, which then gives
\be
\label{eq:Euler2}
\Vv_\x \simeq \frac{e q_\x \, \Ev }{m_\x \, \Gamma_p} + \frac{\gvec}{\Gamma_p} - \frac{D_\x}{n_\x} \, \grad n_\x 
~,
\ee
where $D_\x = T_\x / (m_\x \, \Gamma_p)$ is the diffusion coefficient and we approximated the mCP temperature $T_\x$ as uniform. The first two terms in \Eq{Euler2} correspond to the bulk drift velocity in an external electric or gravitational field, whereas the third term is simply the bulk motion due to thermal diffusion away from pressure gradients. 

\Eq{Euler2} is valid in the limit that mCPs rapidly scatter off of atoms, such that their temperature closely tracks the temperauture $T_N$ of the immediate environment. This requires both the momentum-exchange rate $\Gamma_p$ and the energy-exchange rate $\Gamma_E \simeq (\mu_{\x N} / m_N) \, \Gamma_p$ to be large. In our analysis, we take the incoming mCPs to have an initial temperature $T_\x \simeq T_N$; hence, we have ignored the fact that for the smallest masses and couplings considered in this work, the suppression in $\Gamma_E$ implies that mCPs can gain kinetic energy greater than $T_N$ after accelerating into the accumulator. We have checked that the main effect of such acceleration is to increase the rate for mCP-atomic scattering, since the nuclear charge is less screened by atomic electrons for larger momentum transfer. Our approximation in \Eq{Euler2} is therefore conservative. 

If $e q_\x E \gg m_\x \, g$, then we can ignore the second term in \Eq{Euler2}, in which case the induced current of mCPs is
\be
\label{eq:jxdiff}
\jv_\x  \simeq e q_\x \, \bigg( n_\x  \, \frac{e q_\x \, \Ev }{m_\x \, \Gamma_p} - D_\x \, \grad n_\x  \bigg)
~.
\ee
The standard diffusion equation then follows directly from continuity of this current, $e q_\x \, \dt n_\x + \grad \cdot \jv_\x = 0$. In the steady-state limit, we can ignore the first term involving the time-derivative, which in the case of spherical symmetry yields
\be
\label{eq:diffusion1}
n_\x \, \frac{e q_\x \, \Ev}{m_\x \, \Gamma_p} - D_\x \, \grad n_\x = \frac{\const}{r^2} \, \hat{\rv}
~.
\ee
\Eq{diffusion1} can be analytically solved for the number density $n_\x$ in the simple case that the electric field scales as $\Ev = - E_0 \, (R_0 / r)^{-n} \, \hat{\rv}$ for $r \geq R_0$, where $R_0$ is the radius of the shell and $n$ is some positive integer. In doing so, we normalize the number density to a constant at a distance far from the shell, $\lim_{r \to \infty} n_\x (r) = n_\infty$, and enforce an absorbing boundary condition at the shell, $n_\x (R_0) = 0$, corresponding to a perfectly efficient electric trap. Substituting the solution for $n_\x$ back in \Eq{jxdiff}, we find that for $n=2$ or $n=3$ (corresponding to a monopole or an approximately dipole electric field, respectively),
\be
\label{eq:jxdipole}
\jv_\x (r) = - \frac{(e q_\x)^2 \, n_\infty \, E_0 }{m_\x \, \Gamma_p} \, \bigg( \frac{R_0}{r} \bigg)^2 \hat{\rv} \times
\begin{cases}
\big( 1 - e^{-c_E} \big)^{-1} \simeq 1 & (n=2)
\\
\text{Erf} \big( \sqrt{c_E / 2} \big)^{-1} \, \sqrt{\frac{2}{\pi \, c_E}} \simeq \sqrt{\frac{2}{\pi \, c_E}} & (n=3)
\end{cases}
~~,~~
c_E \equiv \frac{e q_\x \phi_0}{T_\x}
~,
\ee
where in the second equality of each line we took $c_E \equiv e q_\x \phi_0 / T_\x \gg 1$, corresponding to  mCPs that are efficiently bound to the shell. Identifying the accumulation velocity as $V_E = j_\x (R_0) / (e q_\x \, n_\infty)$, we see that \Eq{jxdipole} reproduces \Eq{VE} in the limit that mCPs are highly-collisional and strongly bound to the shell. Note that this applies only to spherically-symmetric potentials; for, e.g., a dipole, we thus expect $\order{1}$ corrections accounting for the fact that the electric field is not radial for an $\order{1}$ fraction of solid angles.

Now, let us consider the ballistic limit, in which case we can ignore collisions between mCPs and air before they accumulate in the potential $\phi_0$ of the driven shell. In this case, we can derive the accumulation of particles for a general spherically-symmetric electric potential. We write the electric potential energy of an mCP as $e q_\x \phi(r) = - \frac{1}{2} \, m_\x \, v_\text{esc}^2 \, |\phi(r) / \phi_0|$, where $\phi_0 = |\phi(R_0)|$ and $v_\text{esc} = \sqrt{2 e q_\x \, \phi_0 / m_\x}$ is the electrical escape velocity of the shell. The trajectory of an mCP with initial velocity $v_\text{th} = \sqrt{3 T_\x / m_\x}$ is determined by the one-dimensional effective potential energy, 
\be
U_\text{eff} (r) = - \frac{1}{2} \, m_\x \, v_\text{esc}^2 \, \bigg| \frac{\phi(r)}{\phi_0} \bigg| + \frac{v_\text{th} \, b}{2 r^2}
~,
\ee
where $b$ is the mCP impact parameter with respect to the center of the potential. At the mCP's distance of closest approach $r_\text{min}$, $U_\text{eff}$ is equal to the mCP's initial kinetic energy, yielding
\be
b = r_\text{min} \, \bigg( 1 + \frac{v_\text{esc}^2}{v_\text{th}^2} \, \bigg| \frac{\phi(r_\text{min})}{\phi_0} \bigg| \, \bigg)^{1/2}
~.
\ee
Taking $r_\text{min} = R_0$ then determines the maximum impact parameter for which an mCP passes through the shell,
\be
b_\text{max} = R_0 \, \bigg( 1 + \frac{v_\text{esc}^2}{v_\text{th}^2} \bigg)^{1/2}
~.
\ee
Thus, for $v_\text{esc} \gg v_\text{th}$, the accumulation rate is dictated by 
\be
v_\text{th}  \, b_\text{max}^2 \simeq v_\text{th}  \, R_0^2 \, \bigg( \frac{v_\text{esc}}{v_\text{th}} \bigg)^2 = R_0^2 \, \frac{e q_\x \, E_0 / m_\x}{v_\text{th} / 2 R_0}
~,
\ee
corresponding to the ballistic regime of \Eq{VE}. Once again, this applies only to a spherically-symmetric potential. In the case of a dipole, which is not symmetric or purely radial, we thus expect that this is not a good approximation when $r_\text{min} \simeq R_0$, in which case an incoming mCP traverses a large range of solid angles before its closest approach to the shell. However, this is not a concern for the majority of the parameter space that we consider since for $r_\text{min} \ll R_0$ mCPs follow approximately radial trajectories, leading to only $\order{1}$ corrections.

\section{Diffusion and Drift}
\label{app:diffusion}

In \Eq{room}, we used that the diffusion flux through a barrier of thickness $\Delta x$ scales as $D_\x / \Delta x$ and that the flux induced by gravity scales as $\Vv_g = \gvec / \Gamma_p$. This can be seen from the fluid equations,
\begin{align}
\label{eq:fluid1}
& \dt \rho_\x + \grad \cdot \jv_\x = 0 
\nl
& \jv_\x \simeq \rho_\x \, \bigg( \frac{e q_\x \, \Ev}{m_\x \, \Gamma_p} + \frac{\gvec}{\Gamma_p} \bigg) - D_\x \, \grad \rho_\x
~,
\end{align}
where $\rho_\x$ and $\jv_\x$ are the millicharge and millicurrent densities. Above, the first line is simply continuity of mCPs. The second line follows from the Euler equation, as in \Eq{jxdiff}, but now including the gravitational term. 

If the gravitational field dominates, then $\jv_\x \simeq \rho_\x \, \Vv_g$, where $\Vv_g = \gvec / \Gamma_p$ is the gravitational drift velocity. We next take the current density $\jv_\x$ in \Eq{fluid1} to be dictated solely by diffusion, $\jv_\x = - D_\x \, \grad \rho_\x$. Define the region of the barrier as $x \in (x_0 , x_\text{abs})$ where $x_\text{abs} - x_0 = \Delta x$ and an electric absorber (trap) is placed at $x_\text{abs}$. We denote the outside air ($x < x_0$) as region 1 and the interior of the barrier as region 2. We also take the diffusion coefficient to be constant in each region. Integrating from $x_1$ in region 1 to $x_2$ in region 2, we then have
\be
\rho_\x(x_2) - \rho_\x(x_1) = -j_\x \, \bigg( \frac{x_0 - x_1}{D_\x (x_1)} + \frac{x_2 - x_0}{D_\x (x_2)} \bigg)
~.
\ee
Enforcing the absorbing boundary condition $n(x_\text{abs}) = 0$, this reduces to
\be
\rho_\x(x) = j_\x \times
\begin{cases}
\frac{x_0 - x}{D_\x (x_1)} + \frac{x_\text{abs} - x_0}{D_\x (x_2)} & (x < x_0)
\\
\frac{x_\text{abs} - x}{D_\x (x_2)} & (x_0 < x < x_\text{abs})
~.
\end{cases}
\ee
Now, we demand that for $x \ll x_0$, $\rho_\x(x) \simeq \bar{\rho}_\x$, where $\bar{\rho}_\x$ is the ambient charge density. Using this in the expression above and solving for $j_\x$ then gives
\be
j_\x \simeq \bar{\rho}_\x \, \bigg( \frac{x_0 }{D_\x (x_1)} + \frac{x_\text{abs} - x_0}{D_\x (x_2)} \bigg)^{-1}
\simeq \bar{\rho}_\x \, \frac{D_\x (x_2)}{\Delta x}
~,
\ee
where in the second equality we took $D_\x (x_2) \ll D_\x (x_1)$ due to the much larger density of the barrier compared to air. 

\end{appendices}

\end{document}

%% file: packages.tex
\usepackage{amsmath,amssymb,amsthm}
\usepackage{graphicx}
\usepackage[usenames, dvipsnames]{xcolor}
\usepackage{slashed}
\usepackage{subfig}
\usepackage{tabularx}
\usepackage{empheq}
\usepackage{mathrsfs}
\usepackage{comment}
\usepackage{bbold}
\usepackage[shortlabels]{enumitem}
\usepackage[normalem]{ulem}
\usepackage{url}
\usepackage{color}

\definecolor{light-gray}{gray}{0.9}
\definecolor{medium-gray}{gray}{0.7}
\definecolor{darkblue}{rgb}{0.0,0.0,0.6}
\definecolor{red}{rgb}{0.9, 0,0}
\definecolor{navy}{rgb}{0.05, 0.05,0.8}
\definecolor{linkcolor}{rgb}{0.0, 0.28, 0.67}
\definecolor{paleblue}{rgb}{0.69, 0.93, 0.93}  

\usepackage[colorlinks]{hyperref}
\hypersetup{
     colorlinks   = true,
     citecolor    = linkcolor,
     linkcolor    = linkcolor,
     urlcolor    = linkcolor
}

\usepackage[titletoc]{appendix}

%% file: macros.tex
\newcommand{\be}{\begin{equation}}
\newcommand{\ee}{\end{equation}}
\newcommand{\nl}{\nonumber \\}
\newcommand{\x}{\chi}

\newcommand{\Ap}{A^\prime}
\newcommand{\mAp}{m_{A^\prime}}
\newcommand{\Lag}{\mathscr{L}}
\newcommand{\eps}{\epsilon}

\newcommand{\p}{\prime}

\newcommand{\grad}{\nabla}

\newcommand{\order}[1]{\mathcal{O}{(#1)}}
\newcommand{\Eq}[1]{Eq.~\ref{eq:#1}}

\newcommand{\Eqs}[2]{Eqs.~\ref{eq:#1} and \ref{eq:#2}} 
\newcommand{\Eqss}[3]{Eqs.~\ref{eq:#1}, \ref{eq:#2}, and  \ref{eq:#3}} 
\newcommand{\Sec}[1]{Sec.~\ref{sec:#1}}
\newcommand{\Secs}[2]{Secs.~\ref{sec:#1} and \ref{sec:#2}} 
\newcommand{\App}[1]{Appendix~\ref{app:#1}}
\newcommand{\Fig}[1]{Fig.~\ref{fig:#1}}
\newcommand{\Figs}[2]{Figs.~\ref{fig:#1} and \ref{fig:#2}}

\newcommand{\bebox}{\begin{empheq}[box=\fcolorbox{light-gray}{light-gray}]{align}}
\newcommand{\eebox}{\end{empheq}}

\newcommand{\DM}{{_\text{DM}}}

\newcommand{\dt}{\partial_t}

\newcommand{\mn}{\mu \nu}

\newcommand{\g}{\gamma}
\newcommand{\const}{\text{constant}}

\newcommand{\jv}{\boldsymbol{j}}

\newcommand{\rv}{{\bf r}}

\newcommand{\Vv}{{\bf V}}

\newcommand{\Ev}{{\bf E}}

\newcommand{\gvec}{\boldsymbol{g}}

\newcommand{\eV}{\text{eV}}
\newcommand{\meV}{\text{meV}}

\newcommand{\MeV}{\text{MeV}}
\newcommand{\GeV}{\text{GeV}}
\newcommand{\TeV}{\text{TeV}}
\newcommand{\V}{\text{V}}

\newcommand{\kV}{\text{kV}}
\newcommand{\MV}{\text{MV}}

\newcommand{\cm}{\text{cm}}
\newcommand{\mm}{\text{mm}}

\newcommand{\m}{\text{m}}
\newcommand{\km}{\text{km}}

\newcommand{\yr}{\text{yr}}

\newcommand{\Uone}{\text{U}(1)}